\documentclass[
    aps,
    reprint,
    amsmath,
    amssymb,
    superscriptaddress
]{revtex4-2}

\usepackage{graphicx}
\usepackage{hyperref}
\usepackage{float}
\hypersetup{
    colorlinks=true,
    linkcolor=blue,
    citecolor=blue,
    urlcolor=blue
}

\DeclareMathOperator{\SiLU}{SiLU}

\begin{document}

\title{NucleoNet: A Nuclear-Aware Determinant Ansatz for Multi-Geometry Neural Variational Monte Carlo}

\author{Qiao Gao}
\affiliation{School of Artificial Intelligence and Data Science, University of Science and Technology of China, Hefei 230026, China}

\author{Yang Huang}
\affiliation{University of Science and Technology of China, Hefei 230026, China}
\affiliation{Suzhou Institute for Advanced Research, University of Science and Technology of China, Suzhou 215213, China}

\author{Enze Hou}
\affiliation{Science, Mathematics and Technology Cluster, Singapore University of Technology and Design, 8 Somapah Road, 487372 Singapore}

\author{Yuzhi Liu}
\email{liuyuzhi17@mails.ucas.ac.cn}
\affiliation{AI for Science Institute, Beijing 100080, China}

\author{Jingrun Chen}
\email{jingrunchen@ustc.edu.cn}
\affiliation{School of Mathematical Sciences, University of Science and Technology of China, Hefei 230026, China}
\affiliation{Suzhou Institute for Advanced Research, University of Science and Technology of China, Suzhou 215213, China}

\author{Han Wang}
\email{wang\_han@iapcm.ac.cn}
\affiliation{National Laboratory of Computational Physics, Institute of Applied Physics and Computational Mathematics, Beijing 100088, China}
\affiliation{HEDPS, CAPT, College of Engineering, Peking University, Beijing, China}

\date{\today}

\begin{abstract}
Neural-network variational Monte Carlo provides an accurate route to solving the electronic Schr\"odinger equation, but many neural wavefunctions are optimized for a single nuclear geometry, requiring separate calculations to map out a potential energy surface.
We introduce NucleoNet, a nuclear-aware determinant ansatz that jointly evolves electronic and nuclear representations through a shared backbone, with antisymmetry enforced by electronic determinants.
Across eight equilibrium small-molecule benchmarks, NucleoNet achieves energies comparable to established neural-wavefunction baselines with substantially fewer trainable parameters, totaling 0.10–0.13 million.
For each molecular system, a single model optimized jointly across geometries describes the N$_2$ dissociation curve, Li displacement in LiCN, and the ethane and formamide minimum-energy paths for bond dissociation.
Models trained on sparse geometry sets generalize to unseen configurations in N$_2$ and LiCN without further optimization.
For N$_2$, training on six geometries yields a mean absolute error (MAE) of $1.13~\mathrm{m}E_h$ on a 32-point evaluation grid, compared with $1.07~\mathrm{m}E_h$ when optimized on all 32 geometries.
Ablation studies on N$_2$ show that propagating nuclear representations improves geometric generalization under both shared and nucleus-specific envelope parameterizations.

\end{abstract}

\maketitle
\section{Introduction}

Neural-network wavefunctions optimized with real-space variational Monte Carlo (VMC) provide expressive representations for solving the electronic Schr\"odinger equation~\citep{carleo2017solving,foulkes2001qmc,needs2020qmc}. 
Architectures such as PauliNet~\citep{hermann2020paulinet}, FermiNet~\citep{pfau2020ferminet}, Psiformer~\citep{glehn2023psiformer}, and LapNet~\citep{li2024lapnet, fu2026empowering, li2025spin} have achieved high accuracy for atoms and molecules. 
Despite the accuracy, neural VMC remains computationally expensive because optimization requires repeated Monte Carlo sampling, local-energy evaluation, wavefunction-gradient estimation, and stochastic parameter updates. 
Recent advances have reduced this computational cost through improved architectures~\citep{gerard2022goldstandard}, stochastic-reconfiguration optimizers such as SPRING~\citep{sorella1998green,sorella2007weak,goldshlager2024spring}, and faster Laplacian evaluation~\citep{li2024lapnet}.
A broader review of neural wavefunctions in ab initio quantum chemistry is provided by \citet{hermann2023review}.

Many chemical applications require an accurate potential energy surface rather than an isolated single-geometry solution.
Independently optimizing each geometry along a bond-dissociation curve or reaction pathway is costly and can yield a noisy surface because stochastic errors are uncorrelated across independently optimized geometries~\citep{scherbela2022weightsharing,scherbela2023budget}.
Recent neural-wavefunction approaches address this challenge by sharing information across nuclear configurations, either through shared parameters or by explicitly conditioning the wavefunction on molecular geometry.
Weight-sharing methods reuse most parameters across geometry-specific models, without explicitly mapping nuclear configurations to wavefunction components~\citep{scherbela2022weightsharing}. 
PESNet uses a graph neural network to generate geometry-dependent wavefunction parameters~\citep{gao2022pesnet}, whereas Globe generates localized molecular-orbital embeddings that reparametrize the wavefunction across molecular structures~\citep{gao2023globe}. 
Transferable atomic-orbital models construct orbital descriptors from localized Hartree--Fock orbitals and combine them with electron embeddings to form the Slater-matrix entries~\citep{scherbela2024transferable}. 
Orbformer uses an Orbital Generator and a message-passing neural network (MPNN) that operate only on nuclei to generate orbital components and nuclear representations, while a separate Electron Transformer processes the sampled electrons~\citep{foster2025orbformer}.
In these approaches, geometry-dependent components are constructed from the nuclear configuration or associated mean-field orbitals, without feedback from the evolving electronic representations.

This one-way conditioning raises a distinct architectural question: should nuclear representations remain independent of the sampled electron configuration, or should electrons and nuclei exchange information throughout the wavefunction backbone? 
We investigate the latter design through \textbf{NucleoNet}, a nuclear-aware, determinant-based neural-wavefunction ansatz for VMC. 
In NucleoNet, electrons and nuclei exchange information through shared group summaries at every layer, making the nuclear representations dependent on both molecular geometry and the sampled electronic configuration.
The shared backbone generates electron-aware nuclear representations within a unified architecture, limiting the parameter overhead of geometry conditioning.
The orbital matrices are constructed solely from the electronic representations, and fermionic antisymmetry is enforced through the determinants of the enveloped Slater matrices.
NucleoNet learns a shared wavefunction across nuclear geometries of a given molecular system.

On equilibrium-geometry benchmarks, NucleoNet achieves variational accuracy comparable to established neural-wavefunction baselines with a substantially smaller parameter count.
When optimized jointly across multiple configurations, a single NucleoNet model produces accurate energy profiles for N$_2$, LiCN, ethane, and formamide, spanning variation along a single nuclear coordinate, diatomic dissociation, and bond cleavage along minimum-energy paths involving concerted motion of the remaining nuclei.
For N$_2$ and LiCN, models optimized on sparse subsets of geometries also generalize to unseen geometries without further optimization, yielding energy profiles close to those from full-geometry joint optimization.
Ablation studies examine nuclear propagation and envelope parameter sharing, showing that the joint electron--nucleus backbone improves sparse-geometry generalization under both envelope parameterizations.

\section{NucleoNet}
\label{sec:nucleonet}

\begin{figure*}[t]
    {\centering
    \includegraphics[width=0.9\textwidth]{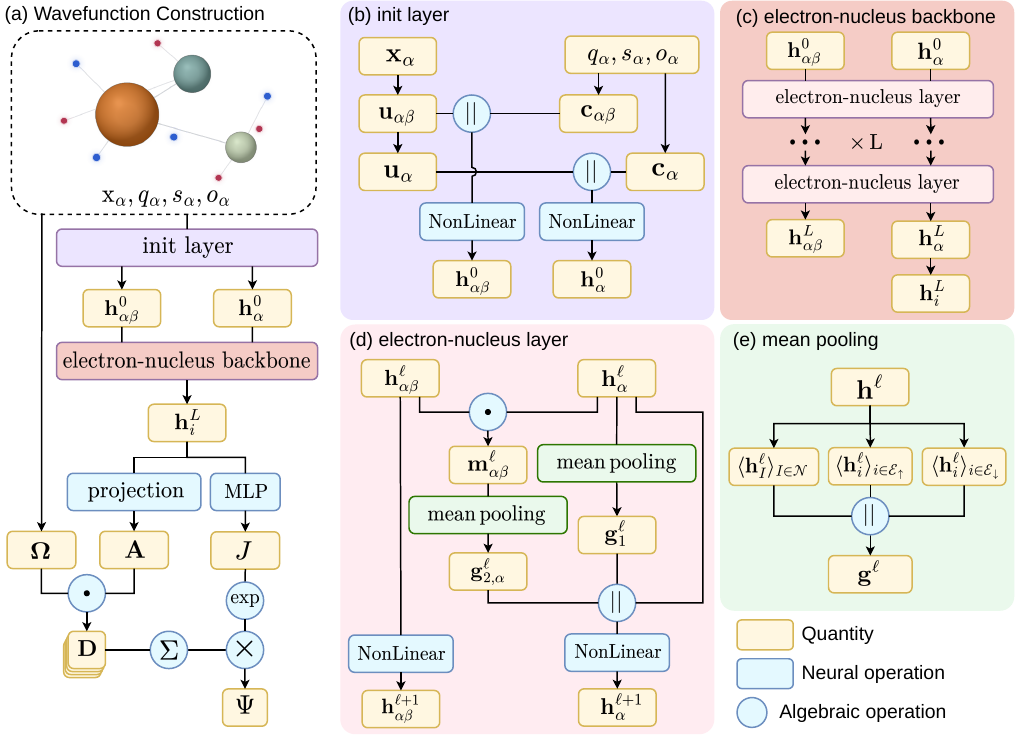}
    \par}
    \caption{
        Overview of NucleoNet.
        (a) Slater--Jastrow wavefunction construction using electronic representations from a shared electron--nucleus backbone.
        (b) Initialization of one- and two-particle representations from geometric features and particle descriptors.
        (c) Shared backbone comprising $L$ electron--nucleus layers.
        (d) One-particle updates using pair-gated messages and group summaries, together with nonlinear two-particle updates.
        (e) Group summaries obtained by separate mean pooling over nuclei, spin-up electrons, and spin-down electrons.
        Symbols $\parallel$, $\cdot$ and $\times$ denote concatenation, elementwise multiplication and scalar multiplication, respectively.
        }
    \label{fig:net}
\end{figure*}

NucleoNet represents electrons and nuclei in a unified particle set and evolves their representations through a shared electron--nucleus backbone.
Figure~\ref{fig:net}(a) summarizes the wavefunction construction.

Throughout this section, we consider a system of $N$ electrons and $M$ nuclei.
We write $\mathbf{r}=\{\mathbf{r}_1,\ldots,\mathbf{r}_N\}$ for electronic configuration and $\mathbf{R}=\{\mathbf{R}_1,\ldots,\mathbf{R}_M\}$ for nuclear configuration, with lowercase indices $i,j$ labeling electrons and uppercase indices $I,J$ labeling nuclei.
The index sets $\mathcal{E}_{\uparrow}$, $\mathcal{E}_{\downarrow}$, and $\mathcal{N}$ collect the spin-up electrons, the spin-down electrons, and the nuclei, with $N_{\uparrow}=|\mathcal{E}_{\uparrow}|$ and $N_{\downarrow}=|\mathcal{E}_{\downarrow}|$ fixed for a given spin state.
The unified particle set treats the two species alike as network inputs, and we write $\mathbf{x}=\{\mathbf{r}_1,\ldots,\mathbf{r}_N,\mathbf{R}_1,\ldots,\mathbf{R}_M\}$ for the combined configuration, with Greek indices $\alpha,\beta$ labeling particles of either type.
For each geometry, the nuclear coordinates $\mathbf R$ are held fixed during electronic Monte Carlo sampling, while they vary across the nuclear configurations considered by a single NucleoNet model.
We write $(\cdot,\cdot)$ for concatenation, $\langle\cdot\rangle$ for permutation-invariant averaging, with a subscript indicating the group averaged over, and $\odot$ for elementwise multiplication.

NucleoNet adopts a Slater--Jastrow form with a multi-determinant expansion,
\begin{equation}
    \Psi(\mathbf{r};\mathbf{R})
    =
    \exp
    \big(
        J(\mathbf{r};\mathbf{R})
        \big)
    \sum_{k=1}^{K}
    D_k(\mathbf{r};\mathbf{R}).
    \label{eq:net_wavefunction}
\end{equation}
Here, $J$ is a neural Jastrow exponent~\citep{jastrow1955manybody,foulkes2001qmc}, $K$ is the number of determinants, and $D_k$ is the determinant of the $k$-th Slater matrix defined in Eq.~\eqref{eq:net_determinant}.
The trainable parameters are left implicit in the notation throughout.

For each ordered particle pair $(\alpha,\beta)$, NucleoNet constructs a pair feature $\mathbf{u}_{\alpha\beta}$ from the logarithmically rescaled relative displacement, together with a corresponding one-particle feature $\mathbf{u}_{\alpha}$:
\begin{equation}
    \begin{aligned}
        \mathbf{u}_{\alpha\beta}
        &=
        \begin{cases}
        \left(
        \dfrac{
            \log(1+\left\| \mathbf{x}_{\alpha\beta} \right\|)
        }{
            \left\| \mathbf{x}_{\alpha\beta} \right\|
        }
        \mathbf{x}_{\alpha\beta},
        \log(1+\left\| \mathbf{x}_{\alpha\beta} \right\|)
        \right),
        & \alpha\neq\beta,
        \\
        \mathbf{0},
        & \alpha=\beta,
        \end{cases}
        \\[3pt]
        \mathbf{u}_{\alpha}
        &=
        \begin{cases}
        \left\langle \mathbf{u}_{\alpha\beta} \right\rangle_{\beta\in
        \mathcal{E}_{\uparrow}\cup\mathcal{E}_{\downarrow}},
        & \alpha\in\mathcal{E}_{\uparrow}\cup\mathcal{E}_{\downarrow},
        \\
        \left\langle \mathbf{u}_{\alpha\beta} \right\rangle_{\beta\in\mathcal{N}},
        & \alpha\in\mathcal{N},
        \end{cases}
    \end{aligned}
    \label{eq:net_pair_and_particle_features}
\end{equation}
where $\mathbf{x}_{\alpha\beta}=\mathbf{x}_\alpha-\mathbf{x}_\beta$.
The logarithmic radial rescaling compresses long-range distances while retaining directional information, following the pair features used in neural wavefunction ansatzes~\citep{pfau2020ferminet,glehn2023psiformer}.
The one-particle feature $\mathbf{u}_\alpha$ averages over particles of the same species as $\alpha$, so that it summarizes the local electronic or nuclear geometry; electron--nucleus geometry is carried by the pair features $\mathbf{u}_{\alpha\beta}$ and enters the one-particle representations through the backbone.

The initial representations are obtained by combining the geometric features with fixed particle and pair descriptors, $\mathbf{c}_\alpha$ and $\mathbf{c}_{\alpha\beta}$, through affine maps followed by sigmoid linear unit (SiLU) activations [Fig.~\ref{fig:net}(b)]:
\begin{equation}
    \begin{aligned}
        \mathbf{h}^0_\alpha
    &=
    \SiLU\big(W^0_1
        (\mathbf{u}_\alpha,
        \mathbf{c}_\alpha)+b^0_1\big),
    \\
    \mathbf{h}^0_{\alpha\beta}
    &=
    \SiLU\big(W^0_2
        (\mathbf{u}_{\alpha\beta},
        \mathbf{c}_{\alpha\beta})+b^0_2\big).
    \end{aligned}
    \label{eq:net_initial_states}
\end{equation}
The descriptors encode particle properties such as normalized particle charge $q_\alpha$, spin $s_\alpha$, and type $\mathbf{o}_\alpha$, together with pairwise combinations of these quantities; the Supplementary Material gives their full definition.
The affine maps are shared across particle types, while particle identity enters through the descriptors and the group-specific pooling operations described below.

The shared backbone consists of $L$ electron--nucleus layers, illustrated in Fig.~\ref{fig:net}(c,d).
At layer $\ell$, NucleoNet maintains one-particle representations $\mathbf{h}^\ell_\alpha\in\mathbb{R}^{h_1}$ and two-particle representations $\mathbf{h}^\ell_{\alpha\beta}\in\mathbb{R}^{h_2}$, where the dimensions $h_1$ and $h_2$ are fixed across layers, and we refer to the layerwise sequence $\mathbf{h}^0_\alpha,\ldots,\mathbf{h}^L_\alpha$ as the stream of particle $\alpha$.
Each layer begins by constructing a directed message $\mathbf{m}^\ell_{\alpha\beta}$ for each ordered pair $(\alpha,\beta)$, by gating the pair representation with the current one-particle representation of the source particle $\beta$:
\begin{equation}
    \mathbf{m}^\ell_{\alpha\beta}
    =
    \mathbf{h}^\ell_{\alpha\beta}
    \odot
    \mathbf{\hat h}^\ell_\beta
    ,
    \label{eq:net_pair_gating}
\end{equation}
where $\mathbf{\hat h}^\ell_\beta$ is the projection of $\mathbf{h}^\ell_\beta$ onto the $h_2$-dimensional two-particle representation space.

NucleoNet then applies permutation-invariant mean pooling over three particle groups (the nuclei $\mathcal{N}$, the spin-up electrons $\mathcal{E}_{\uparrow}$, and the spin-down electrons $\mathcal{E}_{\downarrow}$) and concatenates the results into group summaries, as illustrated in Fig.~\ref{fig:net}(e).
The one-particle summary, shared by all particles, is
\begin{equation}
\begin{aligned}
    \mathbf{g}^\ell_1
    =
    \big(
        \left\langle \mathbf{h}^\ell_I \right\rangle_{I\in\mathcal{N}},
        \left\langle \mathbf{h}^\ell_i \right\rangle_{i\in\mathcal{E}_{\uparrow}},
        \left\langle \mathbf{h}^\ell_i \right\rangle_{i\in\mathcal{E}_{\downarrow}}
    \big),
\end{aligned}
\label{eq:net_global_summary}
\end{equation}
and the pair summary of particle $\alpha$ is
\begin{equation}
\begin{aligned}
    \mathbf{g}^\ell_{2,\alpha}
    =
    \big(
        \left\langle \mathbf{m}^\ell_{\alpha I} \right\rangle_{I\in\mathcal{N}},
        \left\langle \mathbf{m}^\ell_{\alpha i} \right\rangle_{i\in\mathcal{E}_{\uparrow}},
        \left\langle \mathbf{m}^\ell_{\alpha i} \right\rangle_{i\in\mathcal{E}_{\downarrow}}
    \big).
\end{aligned}
\label{eq:net_pair_summary}
\end{equation}
Here $\mathbf{g}^\ell_1$ and $\mathbf{g}^\ell_{2,\alpha}$ couple nuclear and electronic representations in each update, and the averaging renders the backbone equivariant under particle exchange within each group.

The input to the one-particle update is
\begin{equation}
    \mathbf{F}^\ell_\alpha
    =
    \left(
        \mathbf{h}^\ell_\alpha,\,
        \mathbf{g}^\ell_1,\,
        \mathbf{g}^\ell_{2,\alpha}
    \right).
    \label{eq:net_layer_input}
\end{equation}
The one-particle and two-particle representations are then updated as
\begin{equation}
\begin{aligned}
    \mathbf{h}^{\ell+1}_\alpha
    &=
    \frac{1}{\sqrt{2}}
    \Big(
        \mathbf{h}^\ell_\alpha
        +
        \SiLU\big(W^\ell_1\mathbf{F}^\ell_\alpha+b^\ell_1\big)
    \Big),
    \\
    \mathbf{h}^{\ell+1}_{\alpha\beta}
    &=
    \frac{1}{\sqrt{2}}
    \Big(
        \mathbf{h}^\ell_{\alpha\beta}
        +
        \SiLU\big(W^\ell_2\mathbf{h}^\ell_{\alpha\beta}+b^\ell_2\big)
    \Big),
\end{aligned}
\label{eq:net_backbone_update}
\end{equation}
where the factor $1/\sqrt{2}$ rescales the residual sum to help control the magnitude of the representations across layers~\citep{he2016res,gao2023globe}.
The one-particle update draws on both summaries, whereas the two-particle representation evolves through its own nonlinearity.

After $L$ layers, the backbone outputs the final representations $\{\mathbf{h}^L_\alpha\}$.
The electronic representations $\mathbf{h}^L_i$ are used to construct the orbital matrices and the Jastrow factor.

For each determinant $k$, the orbital matrix $A^{k\sigma}\in\mathbb{R}^{N_\sigma\times N}$ is obtained by a linear map of the final electronic representations $\mathbf{h}_i^L$,
\begin{equation}
    A^{k\sigma}_{ia}
    =
    \sum_{c=1}^{h_1}
    W^{k\sigma}_{ac}
    \big(\mathbf{h}^{L}_i\big)_c,
    \quad i\in\mathcal{E}_{\sigma},
    \label{eq:net_orbital_matrix}
\end{equation}
where $\sigma\in\{\uparrow,\downarrow\}$ is the spin index, $a=1,\ldots,N$ is the orbital index, $\big(\mathbf{h}^{L}_i\big)_c$ denotes the $c$-th component of $\mathbf{h}^{L}_i$, and $W^{k\sigma}\in\mathbb{R}^{N\times h_1}$ are trainable weight matrices, with independent parameters for each spin channel.
Row $a$ of $W^{k\sigma}$ generates orbital $a$, and the entry $A^{k\sigma}_{ia}$ is the value of orbital $a$ at electron $i$.

The envelope $\Omega^{k\sigma}_{ia}$ provides an isotropic exponential decay factor for each orbital. We use trainable amplitudes $\pi^{k\sigma}_{a}$ and decay parameters $\lambda^{k\sigma}_{a}$ shared across all nuclei:
\begin{equation}
    \Omega^{k\sigma}_{ia}
    \left(
        \mathbf{r}_i;\mathbf{R}
    \right)
    =
    \pi^{k\sigma}_{a}
    \sum_{I\in\mathcal{N}}
    \exp
    \left(
        -
        \lambda^{k\sigma}_{a}
        \left\|
            \mathbf{r}_i-\mathbf{R}_I
        \right\|
    \right),
\label{eq:net_envelope}
\end{equation}
where $i\in\mathcal{E}_\sigma$ and $a=1,\ldots,N$ indexes the orbitals. 

For each determinant $k$, the orbital matrices $A^{k\sigma}$ and envelopes $\Omega^{k\sigma}$ are combined elementwise and stacked by spin to form the full Slater matrix:
  \begin{equation}
    \Phi_{k}
    \left(
    \mathbf{r};\mathbf{R}
    \right)
    =
    \begin{pmatrix}
    A^{k\uparrow}\odot\Omega^{k\uparrow}
    \\[4pt]
    A^{k\downarrow}\odot\Omega^{k\downarrow}
    \end{pmatrix}.
    \label{eq:net_slater_matrix}
\end{equation}
Each block $A^{k\sigma}\odot\Omega^{k\sigma}$ has $N_\sigma$ rows and $N$ columns, and $\Phi_k$ is an $N\times N$ matrix whose rows run over all electrons and whose columns run over the $N$ orbitals.
The determinant of the $k$-th Slater matrix is then
\begin{equation}
    D_k
    \left(
        \mathbf{r};\mathbf{R}
    \right)
    =
    \det
    \left[
        \Phi_k\left(\mathbf{r};\mathbf{R}\right)
    \right].
    \label{eq:net_determinant}
\end{equation}
The Jastrow exponent is obtained by applying a shared multilayer perceptron (MLP) to the final electronic representations:
\begin{equation}
    J
    \left(
        \mathbf{r};\mathbf{R}
    \right)
    =
    \sum_{i\in\mathcal{E}_{\uparrow}\cup\mathcal{E}_{\downarrow}}
    \mathrm{MLP}_{\mathrm{jas}}
    \left(
        \mathbf{h}^L_i
    \right).
    \label{eq:net_jastrow}
\end{equation}
Here, $\mathrm{MLP}_{\mathrm{jas}}$ returns a scalar.

Substituting the determinants of Eq.~\eqref{eq:net_determinant} and the Jastrow exponent of Eq.~\eqref{eq:net_jastrow} into Eq.~\eqref{eq:net_wavefunction} completes the NucleoNet wavefunction.
The Supplementary Material gives further implementation details.

\section{Results}
\label{sec:results}

We evaluate single-geometry accuracy, joint optimization across geometries, and generalization from sparse geometry sets, followed by ablation studies of nuclear propagation and envelope parameter sharing.
Throughout this section, $s$ denotes the configuration index.
Energies and distances are expressed in atomic units unless otherwise stated, with $E_h$ denoting the hartree and $a_0$ the bohr radius; $1~\mathrm{m}E_h=10^{-3}E_h$.

\subsection{Optimization protocol}
\label{subsec:training_protocol}

For a nuclear configuration $\mathbf{R}_s$, the variational energy of a NucleoNet wavefunction with shared parameters $\theta$ is
\begin{equation}
E_\theta(\mathbf{R}_s)
=
\frac{
\langle
\Psi_\theta(\mathbf{r};\mathbf{R}_s)
|
\hat{H}(\mathbf{R}_s)
|
\Psi_\theta(\mathbf{r};\mathbf{R}_s)
\rangle
}{
\langle
\Psi_\theta(\mathbf{r};\mathbf{R}_s)
|
\Psi_\theta(\mathbf{r};\mathbf{R}_s)
\rangle
}.
\end{equation}
Here, $\hat H(\mathbf R_s)$ is the fixed-nuclei Born--Oppenheimer Hamiltonian, including the nuclear repulsion energy.
For multi-geometry optimization, a single parameter vector $\theta$ is shared across all training configurations $\{\mathbf{R}_s\}_{s=1}^{N_{\mathrm{train}}}$.
We optimize the equally weighted mean variational energy
\begin{equation}
\mathcal{L}(\theta)
=
\frac{1}{N_{\mathrm{train}}}
\sum_{s=1}^{N_{\mathrm{train}}}
E_\theta(\mathbf{R}_s).
\label{eq:multigeometry_objective}
\end{equation}
Electronic configurations are sampled separately for each nuclear geometry from the corresponding wavefunction distribution.

The variational optimization stage uses SPRING~\citep{goldshlager2024spring}, an optimizer based on stochastic reconfiguration (SR), which is closely related to natural-gradient optimization in VMC~\citep{sorella1998green,sorella2007weak}.
SPRING reuses information from recent minibatches while avoiding the explicit construction and inversion of the full SR matrix.
All reported NucleoNet experiments are performed on a single NVIDIA A100 GPU.
Detailed optimization protocols and hyperparameters are provided in the Supplementary Material.

\begin{figure}[t]
    {\centering
    \includegraphics[width=0.48\textwidth]{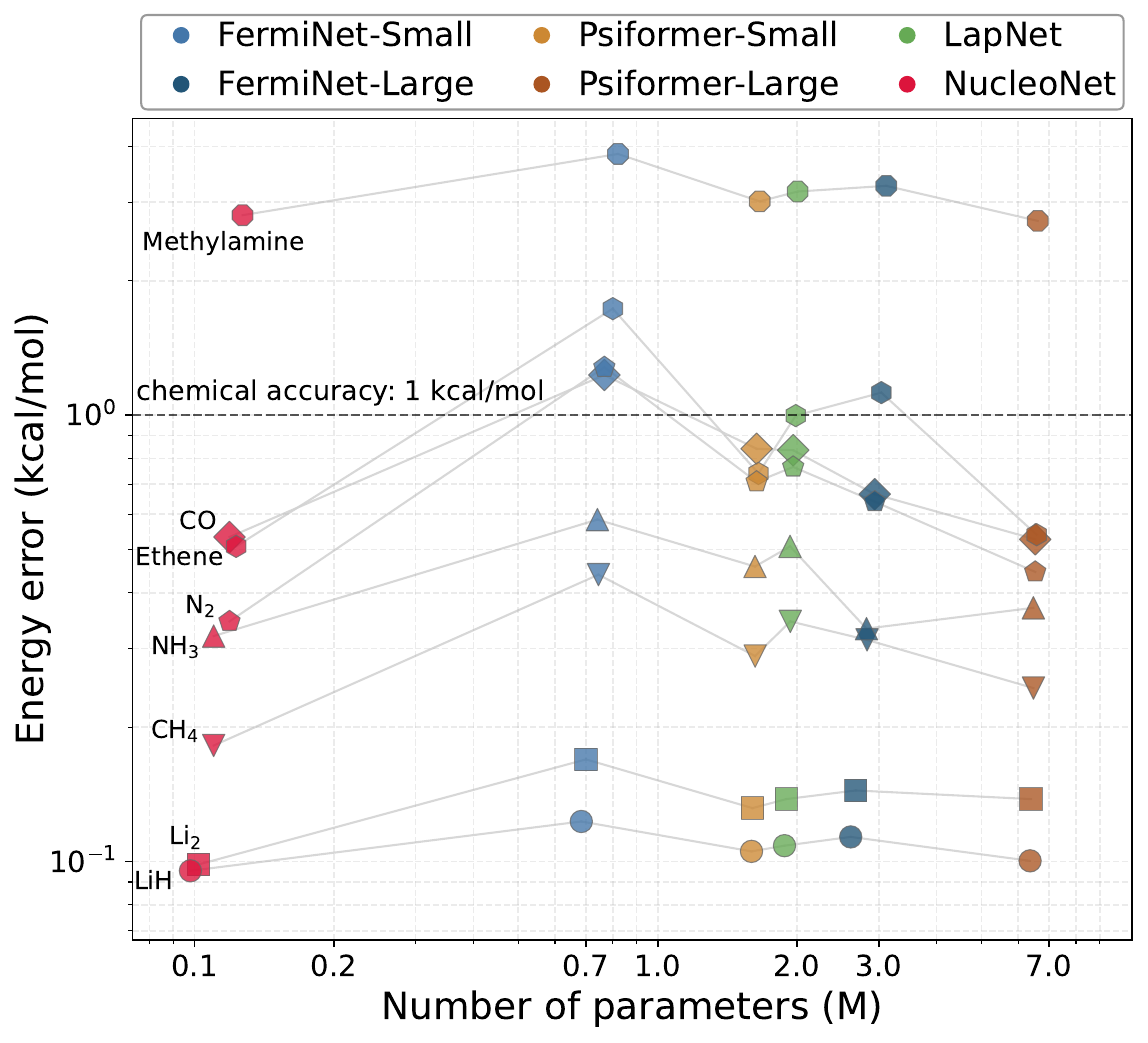}
    \par}
    \caption{
        Energy error versus number of trainable parameters for equilibrium small-molecule benchmarks.
        The lower-left region corresponds to lower errors and fewer parameters.
        Results for the same molecule use the same marker style and are connected by gray lines as a visual guide.
        NucleoNet lies in the low-parameter, low-error region relative to FermiNet, Psiformer, and LapNet baselines.
        }
    \label{fig:params_plot}
\end{figure}

\subsection{Single-geometry benchmarks}
\label{subsec:small_molecules}

We first evaluate the accuracy and parameter efficiency of NucleoNet on eight equilibrium small-molecule benchmarks, using geometries from previous neural VMC studies~\citep{pfau2020ferminet,glehn2023psiformer,li2024lapnet}.
The FermiNet and Psiformer results are taken from Ref.~\citep{glehn2023psiformer}, and the LapNet results from Ref.~\citep{li2024lapnet}.
Energy errors are measured relative to coupled-cluster singles and doubles with perturbative triples [CCSD(T)] energies extrapolated to the complete-basis-set (CBS) limit, as reported in the FermiNet study~\citep{pfau2020ferminet}.
Since VMC provides a variational upper bound to the ground-state energy, lower optimized energies indicate a better variational approximation, up to statistical uncertainty~\citep{foulkes2001qmc,helgaker2014molecular}.

As shown in Fig.~\ref{fig:params_plot}, NucleoNet achieves variational energies comparable to established neural-wavefunction baselines across the eight systems, while using substantially fewer parameters.
NucleoNet has about $0.10$--$0.13$ million trainable parameters, compared with several hundred thousand to multiple million for FermiNet, Psiformer, and LapNet.
Detailed energies, statistical uncertainties and parameter counts are provided in the Supplementary Material.

\subsection{Nitrogen dissociation curve}
\label{subsec:n2_curve}

\begin{figure}[t]
    {\centering
    \includegraphics[width=\columnwidth]{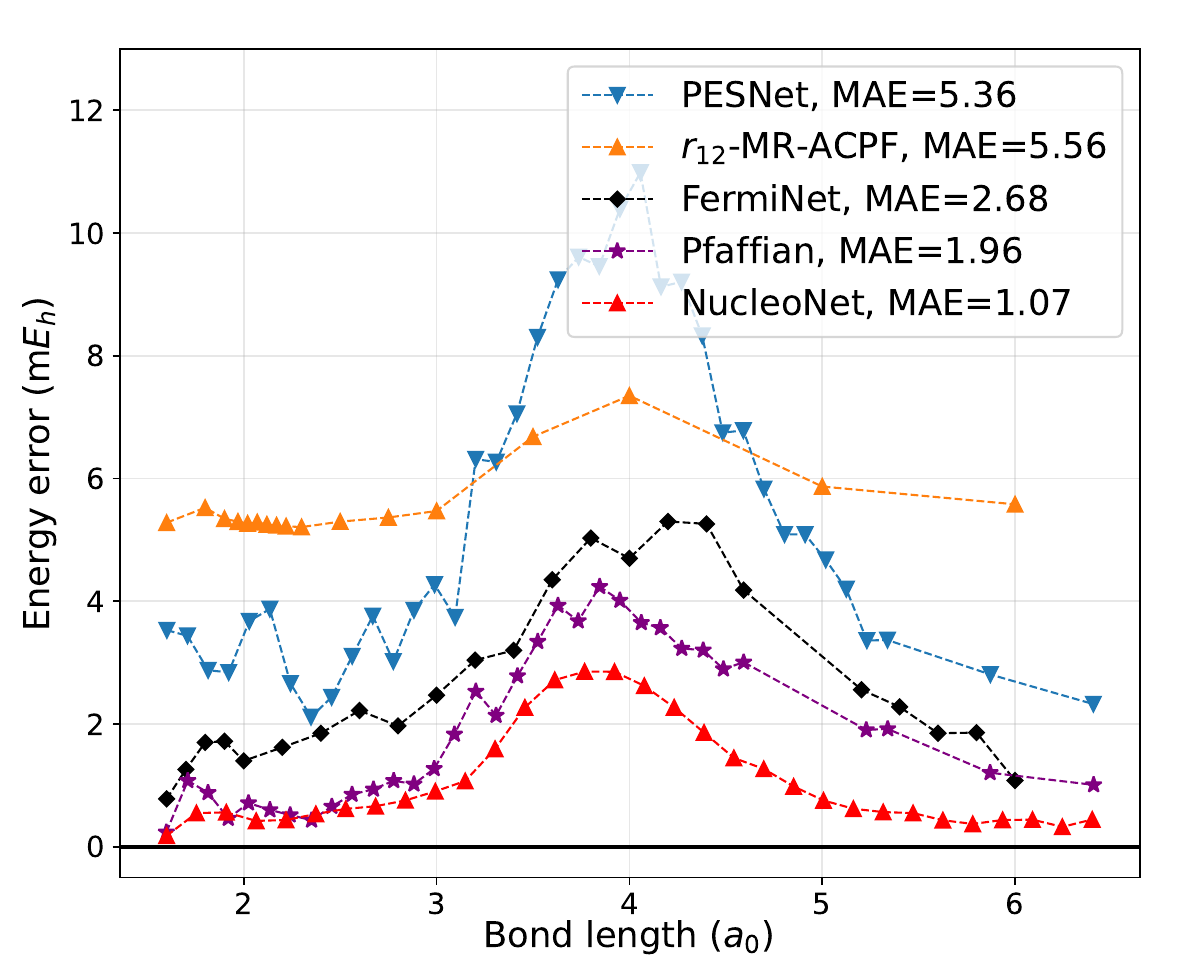}
    \par}
    \caption{
            Energy errors along the N$_2$ dissociation curve relative to the MLR$_4(6,8)$ reference.
            A single NucleoNet model is jointly optimized on 32 bond lengths uniformly spaced between $1.6$ and $6.4~a_0$.
            The MAEs in the legend are computed over the available points for each method.
            }
    \label{fig:n2_dissociation}
\end{figure}

We next evaluate NucleoNet on the dissociation curve of N$_2$, a standard benchmark for methods treating strong static correlation.
As the N$\equiv$N triple bond is stretched, the electronic structure develops strong multireference character, making single-reference descriptions increasingly unreliable~\citep{lyakh2012multireference}.
A single NucleoNet model is optimized jointly on 32 bond lengths uniformly spaced between $1.6$ and $6.4~a_0$ and evaluated on the same grid, which spans the compressed repulsive region, equilibrium, and the multireference regime of the stretched bond.

We quantify the deviation from the reference curve using the mean absolute error (MAE):
\begin{equation}
    \mathrm{MAE}
    =
    \frac{1}{N_L}
    \sum_{s=1}^{N_L}
    \left|
        E_{\mathrm{method}}(L_s)
        -
        E_{\mathrm{ref}}(L_s)
    \right|.
\label{eq:n2_mae}
\end{equation}
Here, $L_s$ is the bond length at the $s$-th evaluated geometry, and $N_L$ is the number of geometries used for the corresponding method.
The reference $E_{\mathrm{ref}}$ is the Morse/long-range (MLR) potential MLR$_4(6,8)$ of Ref.~\citep{leroy2006accurate}, placed on an absolute-energy scale by anchoring its dissociation limit to twice the atomic nitrogen energy per Ref.~\citep{PhysRevA.47.3649}.

As shown in Fig.~\ref{fig:n2_dissociation}, NucleoNet achieves an MAE of $1.07~\mathrm{m}E_h$, the smallest among the compared methods.
For comparison, we obtain MAEs of $2.68$ and $1.96~\mathrm{m}E_h$ for FermiNet and Pfaffian, respectively, using the curves from Ref.~\citep{gao2024neuralpfaffians}.
The PESNet curve from Ref.~\citep{gao2022pesnet} and the tabulated $r_{12}$-MR-ACPF data from Ref.~\citep{robert1998r12} give MAEs of $5.36$ and $5.56~\mathrm{m}E_h$, respectively.

The NucleoNet error profile is smooth across the dissociation path, with the largest deviations from the MLR$_4(6,8)$ reference in the intermediate bond-stretching region. 
The error decreases toward dissociation, and NucleoNet remains closer to the reference than the other methods over most of the curve.

\begin{figure}[!ht]
    {\centering
    \includegraphics[width=0.47\textwidth]{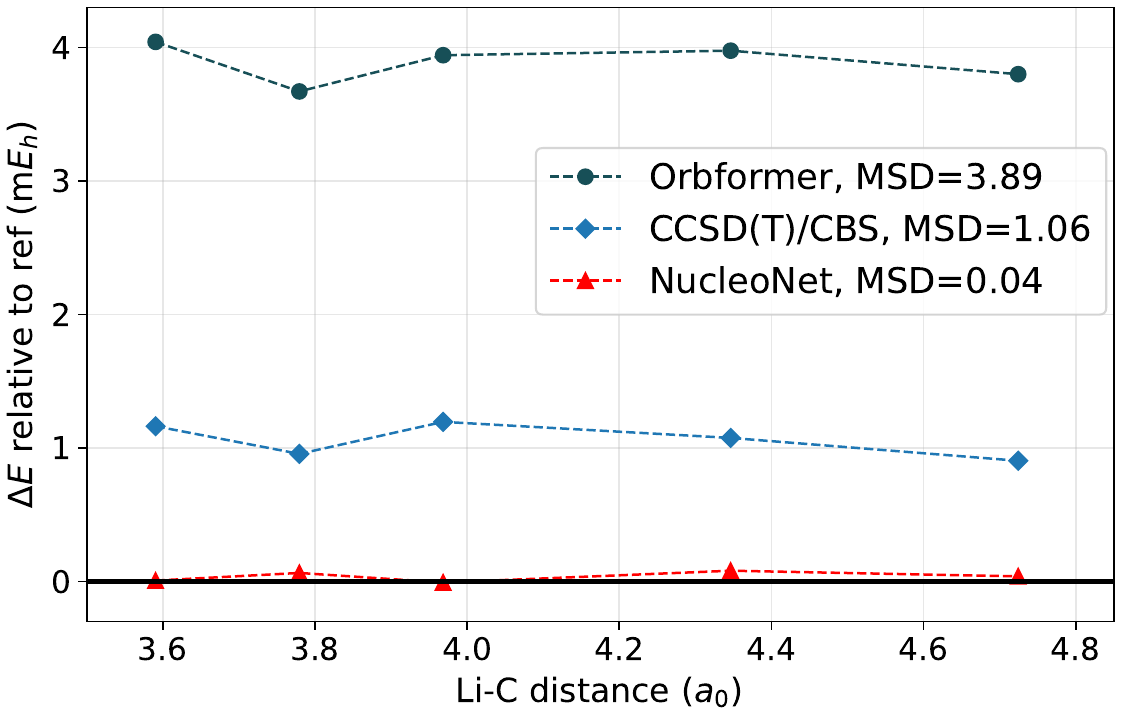}
    \par}
    \caption{
    Signed energy differences across the five LiCN geometries as a function of the Li--C distance, relative to independently optimized single-geometry NucleoNet energies.
    The Orbformer results correspond to the model fine-tuned from the LAC checkpoint; the Orbformer and CCSD(T)/CBS energies are taken from Ref.~\citep{foster2025orbformer}.    
    The black zero line denotes the single-geometry NucleoNet baseline, and the MSD values in the legend are averaged over the five geometries.
    }
    \label{fig:licn_curve}
\end{figure}

\begin{figure*}[t]
    {\centering
    \includegraphics[width=0.95\textwidth]{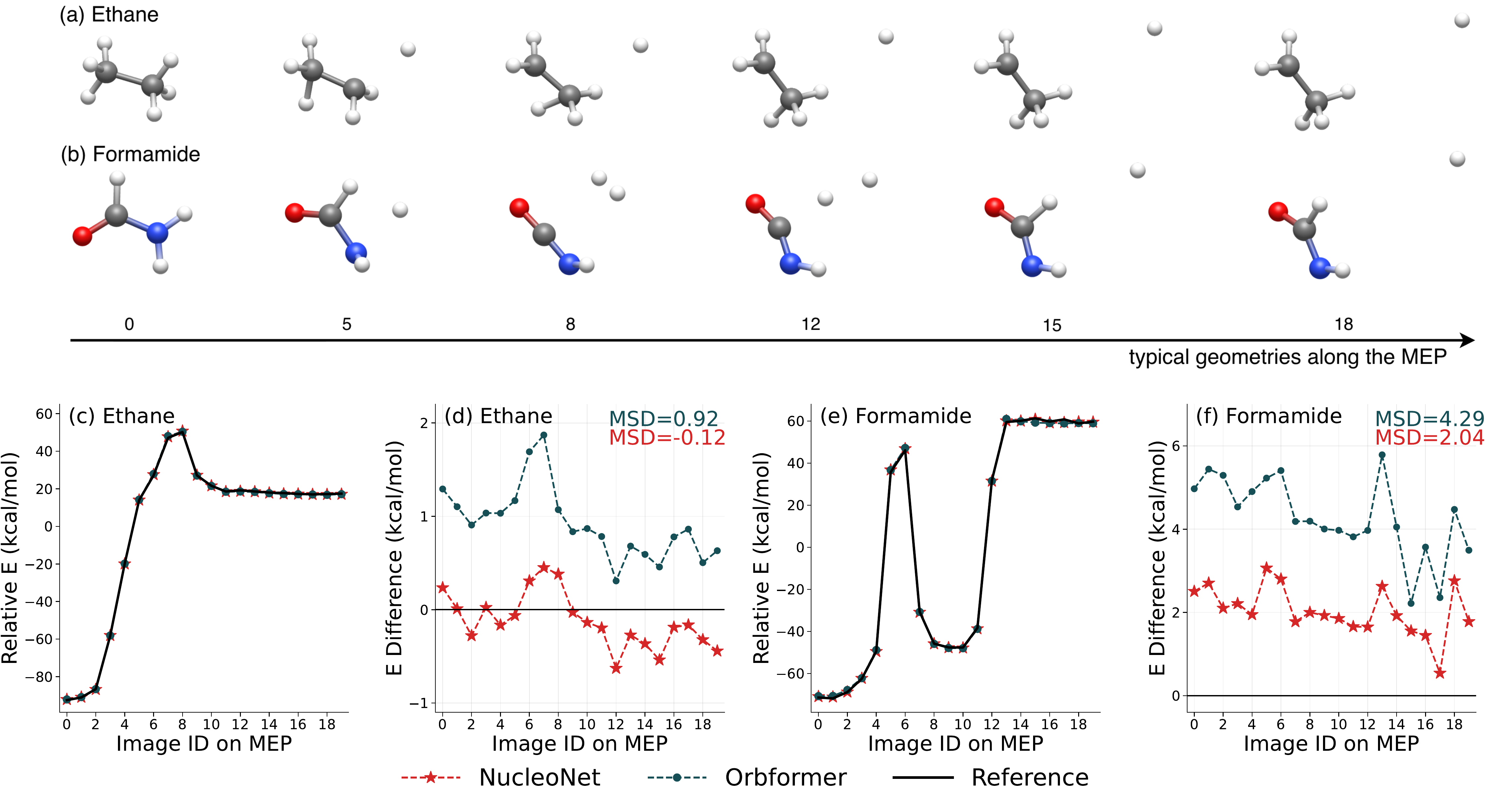}
    \par}
    \caption{
    Multi-geometry dissociation benchmarks for ethane and formamide.
    Panels (a) and (b) show representative molecular geometries along the corresponding minimum-energy paths (MEPs).
    Panels (c) and (e) show mean-centered relative energies, while panels (d) and (f) show signed energy differences relative to the reference calculations reported in Ref.~\citep{foster2025orbformer}.
    The MSD values are averaged over all geometries along each path and energies are reported in kcal/mol ($1~\mathrm{kcal/mol}=1.594~\mathrm{m}E_h$).
    }
    \label{fig:ethane_formamide}
\end{figure*}

\subsection{Li displacement in LiCN}
\label{subsec:licn_curve}

We next consider a controlled multi-geometry setting based on the LiCN benchmark used in Orbformer~\citep{foster2025orbformer}. 
The C and N nuclei remain fixed, while the Li nucleus is displaced along the molecular axis over Li--C distances of approximately $3.6$ to $4.7~a_0$.
This construction isolates the response of the wavefunction to the motion of the Li center relative to the fixed CN fragment.

A single set of NucleoNet model parameters is optimized jointly across all geometries.
To assess the effect of sharing parameters across geometries, we compare the jointly optimized model with independently optimized single-geometry NucleoNet calculations:
\begin{equation}
    \Delta E(s)
    =
    E_{\mathrm{method}}(s)
    -
    E_{\mathrm{NucleoNet}}^{\mathrm{single}}(s),
\end{equation}
and summarize the overall signed energy shift using the mean signed difference (MSD),
\begin{equation}
    \mathrm{MSD}
    =
    \frac{1}{N_{\mathrm{geom}}}
    \sum_{s=1}^{N_{\mathrm{geom}}}
    \Delta E(s),
    \label{eq:msd}
\end{equation}
where $N_{\mathrm{geom}}=5$.
Negative values indicate energies below the independently optimized NucleoNet baseline.

As shown in Fig.~\ref{fig:licn_curve}, jointly optimized NucleoNet energies remain close to the independently optimized single-geometry results, with an MSD of $+0.04~\mathrm{m}E_h$.
CCSD(T)/CBS and Orbformer yield MSDs of $+1.06$ and $+3.89~\mathrm{m}E_h$, respectively.
NucleoNet reaches lower energies than both methods at every geometry.

\subsection{Ethane and formamide dissociation pathways}
\label{subsec:ethane_curve}

We next evaluate NucleoNet on the ethane and formamide minimum-energy paths (MEPs) reported in the Orbformer benchmark~\citep{foster2025orbformer}. 
These paths describe homolytic C--H cleavage, $\mathrm{CH_3CH_3}\to\mathrm{CH_3CH_2}\cdot+\mathrm{H}\cdot$, and homolytic N--H cleavage, $\mathrm{HCONH_2}\to\mathrm{HCONH}\cdot+\mathrm{H}\cdot$. 
Each path contains 20 nudged-elastic-band~\citep{Nudged1998} images, with several nuclear coordinates changing as the bond breaks. 
We jointly optimize a separate NucleoNet model across the images for each molecule.

The reference energies are independent single-geometry Psiformer VMC calculations reported in Ref.~\citep{foster2025orbformer}. 
We quantify the signed energy difference as $\Delta E(s)=E_{\mathrm{method}}(s)-E_{\mathrm{ref}}(s)$ and report its mean, the MSD defined in Eq.~\eqref{eq:msd}, with $N_{\mathrm{geom}}=20$. 
Negative values indicate energies below the reference. 

As shown in Fig.~\ref{fig:ethane_formamide}, NucleoNet stays within approximately $0.6~\mathrm{kcal/mol}$ of the reference at every ethane image and has an MSD of $-0.12~\mathrm{kcal/mol}$, compared with $0.92~\mathrm{kcal/mol}$ for Orbformer.
For formamide, NucleoNet lies above the reference at every image, with an MSD of $2.04~\mathrm{kcal/mol}$, compared with an MSD of $4.29~\mathrm{kcal/mol}$ for Orbformer.
For both systems, NucleoNet reproduces the shape of the reference energy profile and yields lower total energies than Orbformer at every image.

\begin{figure}[t]
    {\centering
    \includegraphics[width=0.47\textwidth]{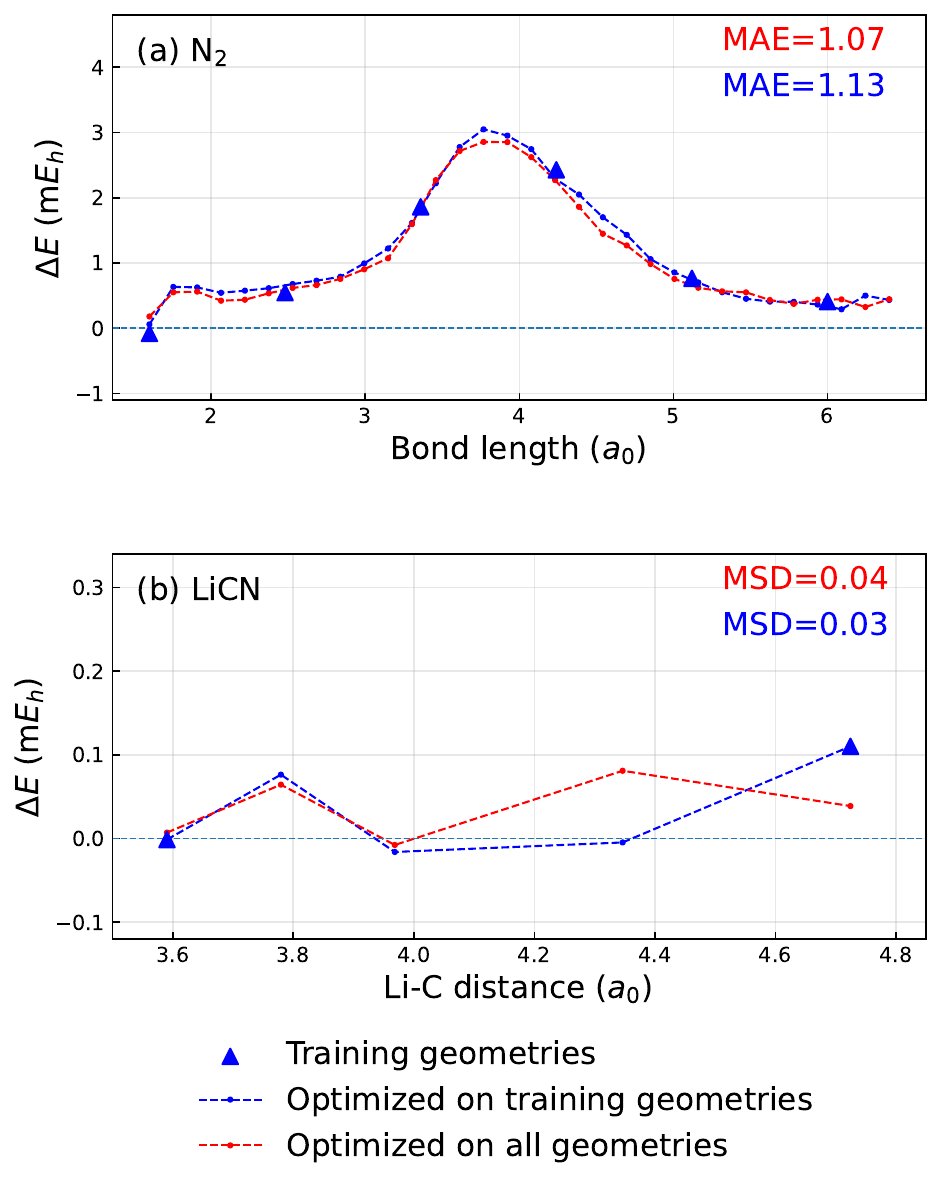}
    \par}
    \caption{
    Generalization from sparse geometry sets.
    Blue triangles mark the training geometries, blue curves show the model optimized on the training geometries and evaluated at all geometries, and red curves show full-geometry joint optimization.
    The upper-right values give the corresponding benchmark metric (MAE for N$_2$ and MSD for LiCN), using the same references and units as in the preceding sections.
    For N$_2$, six uniformly spaced bond lengths are used for optimization; for LiCN, only the two endpoint geometries are used.
    }
    \label{fig:geometry_Generalization}
\end{figure}

\subsection{Generalization from sparse geometry sets}
\label{subsec:geometry_Generalization}

We next test whether a NucleoNet model optimized on a sparse set of geometries generalizes to the remaining geometries without further optimization.
Fig.~\ref{fig:geometry_Generalization} summarizes the results for N$_2$ and LiCN.
The reference energies and metrics are the same as in the preceding benchmarks.

For N$_2$, the model is optimized on six bond lengths uniformly spaced between $1.6$ and $6.0~a_0$ and evaluated on 32 bond lengths uniformly spaced between $1.6$ and $6.4~a_0$, consistent with Sec.~\ref{subsec:n2_curve}.
Training on six geometries yields an evaluation MAE of $1.13~\mathrm{m}E_h$, close to the $1.07~\mathrm{m}E_h$ obtained by joint optimization on all 32 geometries.

For LiCN, only the two endpoint geometries are included in the optimization.
The three intermediate configurations are evaluated without further optimization, with an MSD of $0.03~\mathrm{m}E_h$ evaluated over all five geometries, compared with $0.04~\mathrm{m}E_h$ for full-geometry joint optimization.

\subsection{Ablation studies}
\label{subsec:ablation}

We examine the effects of envelope parameter sharing and the propagation of nuclear representations through two modifications of NucleoNet.
The suffix ``-NS'' denotes replacing the nucleus-shared envelope coefficients of NucleoNet with nucleus-specific coefficients, following the isotropic envelope parameterization of FermiNet~\citep{spencer2020better}.
The suffix ``-EB'' denotes an electron-only backbone that retains nuclear geometry through pooled initialization features but does not propagate nuclear representations.
Combining these modifications gives four models: NucleoNet, NucleoNet-NS, NucleoNet-EB, and NucleoNet-EB-NS.
Detailed definitions are provided in the Supplementary Material.

We compare all four models at the equilibrium geometries of NH$_3$ and N$_2$ and on the N$_2$ sparse-geometry benchmark.
The latter uses the six training geometries and the 32-point evaluation grid of Sec.~\ref{subsec:geometry_Generalization}.
All errors are measured against the reference energies used in the corresponding benchmarks above.
Table~\ref{tab:ablation} reports the single-geometry energy errors, $\Delta E=E_{\mathrm{model}}-E_{\mathrm{ref}}$, and the N$_2$ training and evaluation MAEs.

\begin{figure}[t]
\centering
\includegraphics[width=\columnwidth]{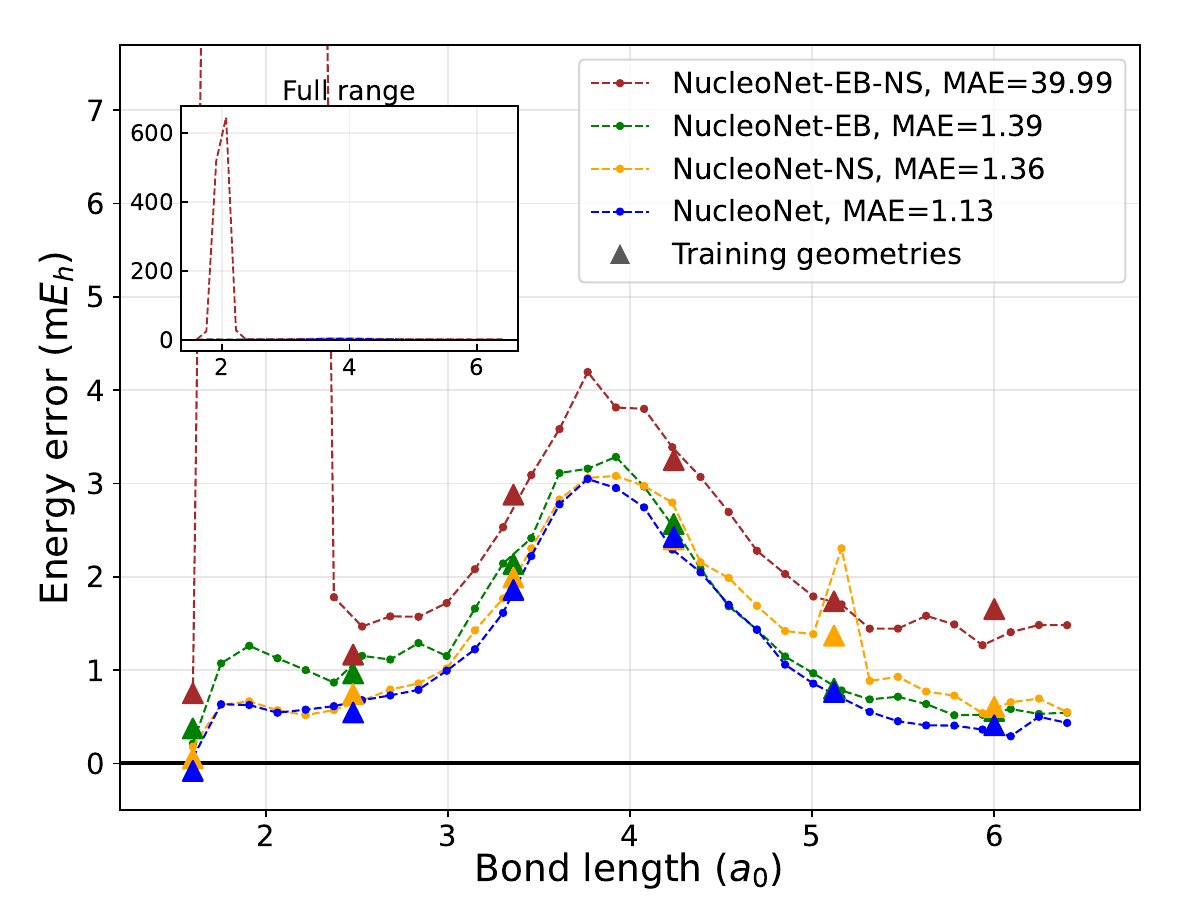}
\caption{
Energy errors along the N$_2$ dissociation curve for the four model variants relative to the MLR$_4(6,8)$ reference.
Circles connected by dashed lines show results on the evaluation grid, and triangles mark results at the six training geometries.
The legend reports the evaluation MAE in $\mathrm{m}E_h$.
The inset shows the full error range, including the large errors of NucleoNet-EB-NS near $2~a_0$.
}
\label{fig:ablation_n2_error}
\end{figure}

\begin{table*}[t]
\centering
\caption{
Ablation results for the NH$_3$ and N$_2$ single-geometry benchmarks and the N$_2$ sparse-geometry benchmark.
The N$_2$ training and evaluation MAEs are computed over the six training geometries and the 32-point evaluation grid, respectively.
All errors use the corresponding benchmark references and are reported in $\mathrm{m}E_h$.
}
\label{tab:ablation}
\setlength{\tabcolsep}{8pt}
\begin{tabular}{lcccc}
\hline
Model & NH$_3$ $\Delta E$ & N$_2$ $\Delta E$ & N$_2$ training MAE & N$_2$ evaluation MAE \\
\hline
NucleoNet       & 0.51 & 0.55 & 1.01 & 1.13 \\
NucleoNet-NS    & 0.52 & 0.66 & 1.19 & 1.36 \\
NucleoNet-EB    & 0.56 & 0.73 & 1.23 & 1.39 \\
NucleoNet-EB-NS & 0.56 & 0.77 & 1.91 & 39.99 \\
\hline
\end{tabular}
\end{table*}

At equilibrium, the four models have similar energy errors, spanning $0.51$--$0.56~\mathrm{m}E_h$ for NH$_3$ and $0.55$--$0.77~\mathrm{m}E_h$ for N$_2$.
The differences are much larger in the sparse-geometry N$_2$ benchmark.
NucleoNet-EB-NS has an evaluation MAE of $39.99~\mathrm{m}E_h$ despite a training MAE of $1.91~\mathrm{m}E_h$, whereas the other three models retain evaluation MAEs close to their training values.
Figure~\ref{fig:ablation_n2_error} shows localized error spikes in both nucleus-specific variants: a large deviation near $2~a_0$ for NucleoNet-EB-NS, between the first two training geometries, and a smaller peak near $5.16~a_0$ for NucleoNet-NS.
The shared-envelope variants show smoother error profiles.

Propagating nuclear representations lowers the evaluation MAE under both envelope parameterizations: from $1.39$ to $1.13~\mathrm{m}E_h$ with shared coefficients, and from $39.99$ to $1.36~\mathrm{m}E_h$ with nucleus-specific coefficients.
Sharing envelope coefficients also improves evaluation accuracy under both backbones, and NucleoNet achieves the lowest errors across all ablation benchmarks.
These comparisons show that the joint electron--nucleus backbone improves geometric generalization and reduces sensitivity to the envelope parameterization.

\section{Discussion}
\label{sec:discussion}

NucleoNet treats nuclei as participants in the wavefunction backbone, allowing nuclear and electronic representations to evolve jointly.
The results show that this architecture combines competitive single-geometry accuracy with a compact parameterization and supports a shared wavefunction across geometries of a molecular system.
The multi-geometry benchmarks cover diatomic dissociation, displacement of an atomic center, and bond cleavage involving concerted nuclear motion.
Sparse-geometry training yields energy profiles close to those from full-geometry optimization for N$_2$ and LiCN.
The N$_2$ ablations show that propagating nuclear representations improves generalization under both envelope choices and reduces sensitivity to the envelope parameterization.
Sharing envelope coefficients provides a further improvement, with the complete model achieving the lowest errors across the ablation benchmarks.

Extending sparse-geometry generalization to higher-dimensional nuclear configurations and transferring across molecular systems are natural next steps.
The all-particle pair interactions scale quadratically with particle number, motivating sparse or locality-based interactions for larger systems.
Future work will test whether learned or symmetry-aware descriptors improve generalization across nuclear configurations.

\section*{Supporting Information}

Network implementation details, optimization and sampling settings,
molecular benchmark energies, sparse-geometry evaluation results,
and ablation data (PDF).

\section*{Data and Software Availability}

The numerical results are reported in the article and its Supporting Information. 
The NucleoNet source code, JSON input files containing molecular geometries and calculation settings, and machine-readable energy summaries with statistical uncertainties are available at \url{https://github.com/Qiao395369/NucleoNet}. 
The repository also provides the Python scripts used to plot the numerical results and instructions for running the calculations and regenerating the plots.

\section*{Acknowledgments}

The work is supported by the National Key R\&D Program of China (Grant No. 2022YFA1004300) and the National Natural Science Foundation of China (Grant Nos. 12425113, 12525113 and 12561160120).
The authors thank Mengsa Wang and Yifei Sun for helpful discussions on optimization, and Xuelan Wen for helpful discussions on N$_2$ benchmarks.

\bibliography{references}

\end{document}


\onecolumngrid

\title{Supplementary Material}
\maketitle

Unless otherwise stated, energies and distances are reported in hartree ($E_h$) and bohr ($a_0$), respectively.
Values in parentheses indicate statistical uncertainties in the last reported digits.
For NucleoNet results, these uncertainties represent autocorrelation-corrected standard errors of the mean energy.
Uncertainties accompanying literature results are reproduced as reported in the original publications.

\section{Network Implementation Details}
\label{app:network_details}
This section details the implementation of NucleoNet.

\subsection{Multilayer Perceptron}
\label{app:residual_nonlinear_blocks}

A multilayer perceptron (MLP) is constructed as the composition of multiple nonlinear blocks:
\begin{equation}
    \mathrm{MLP}(x)
    =
    W_{\mathrm{out}}\bigl(
    \mathrm{NonLinear}_{L_{\rm MLP}}
    \circ\cdots\circ
    \mathrm{NonLinear}_{1}(x)\bigr)
    + b_{\mathrm{out}}.
\end{equation}
Each nonlinear block applies an affine transformation followed by an activation function $\varphi$, with residual mixing when the input and output dimensions agree.
\begin{equation}
\mathrm{NonLinear}_{\ell}(x)
=
\begin{cases}
    \dfrac{x+\varphi(W_\ell x+b_\ell)}{\sqrt{2}},
    & d_{\mathrm{in}}=d_{\mathrm{out}},\\[6pt]
    \varphi(W_\ell x+b_\ell),
    & d_{\mathrm{in}}\ne d_{\mathrm{out}}.
\end{cases}
\end{equation}

\subsection{Particle and Pair Descriptors}
\label{app:particle_pair_descriptors}

NucleoNet encodes particle-type information via descriptor features to represent spin-up electrons, spin-down electrons, and nuclei.
Let $\mathbf{o}_\alpha$ denote the raw particle-type one-hot vector, $q_\alpha$ the normalized particle charge, and $s_\alpha$ the spin marker.
The charge is normalized with respect to the maximum nuclear charge $Z_{\max}$:
\begin{equation}
    q_\alpha
    =
    \begin{cases}
    -1/Z_{\max},
    &
    \alpha\in\mathcal{E}_{\uparrow}\cup\mathcal{E}_{\downarrow},
    \\[4pt]
    Z_I/Z_{\max},
    &
    \alpha=I\in\mathcal{N}.
    \end{cases}
    \label{eq:app_charge_norm}
\end{equation}

The spin marker is
\begin{equation}
    s_\alpha
    =
    \begin{cases}
    +1,
    &
    \alpha\in\mathcal{E}_{\uparrow},
    \\[4pt]
    -1,
    &
    \alpha\in\mathcal{E}_{\downarrow},
    \\[4pt]
    0,
    &
    \alpha\in\mathcal{N}.
    \end{cases}
    \label{eq:app_spin_marker}
\end{equation}

The particle descriptor is
\begin{equation}
    \mathbf{c}_\alpha
    =
    \operatorname{concat}
    \left(
        \mathbf{o}_\alpha,\,
        q_\alpha,\,
        |q_\alpha|,\,
        q_\alpha^2,\,
        \operatorname{sign}\!\left(q_\alpha\right),\,
        s_\alpha
    \right).
    \label{eq:app_particle_descriptor}
\end{equation}

The pair descriptor is
\begin{equation}
\begin{aligned}
    \mathbf{c}_{\alpha\beta}
    =
    \operatorname{concat}
    \left(
        \mathbf{o}_\alpha,\,
        \mathbf{o}_\beta,\,
        q_\alpha,\,
        q_\beta,\,
        q_\alpha q_\beta,\,
        |q_\alpha q_\beta|,\,
        s_\alpha,\,
        s_\beta,\,
        s_\alpha s_\beta
    \right).
\end{aligned}
\label{eq:app_pair_descriptor}
\end{equation}

These descriptors are concatenated with the corresponding geometric features before initialization of the one-particle and two-particle representations.

\subsection{Nuclear-Aware Shared Backbone}
\label{app:shared_backbone}

At each layer, NucleoNet constructs directed pair-gated messages from the one-particle and two-particle representations.
When the one-particle and two-particle representation dimensions differ, the one-particle representation is first projected to the two-particle representation dimension:
\begin{equation}
    \mathbf{\hat h}^\ell_\alpha
    =
    W^\ell \mathbf{h}^\ell_\alpha + b^\ell.
    \label{eq:app_projection}
\end{equation}

For each ordered particle pair $(\alpha,\beta)$, the pair-gated message is
\begin{equation}
    \mathbf{m}^\ell_{\alpha\beta}
    =
    \mathbf{h}^\ell_{\alpha\beta}
    \odot
    \mathbf{\hat h}^\ell_\beta ,
    \label{eq:app_pair_gating}
\end{equation}
where $\beta$ denotes the source particle.
Thus, the message associated with a pair depends on both the two-particle representation and the current one-particle representation of the source particle.

The permutation-invariant one-particle group summary is
\begin{equation}
\begin{aligned}
    \mathbf{g}^\ell_1
    =
    \operatorname{concat}
    \left(
        \left\langle \mathbf{h}^\ell_I \right\rangle_{I\in\mathcal{N}},\,
        \left\langle \mathbf{h}^\ell_i \right\rangle_{i\in\mathcal{E}_{\uparrow}},\,
        \left\langle \mathbf{h}^\ell_i \right\rangle_{i\in\mathcal{E}_{\downarrow}}
    \right),
\end{aligned}
\label{eq:app_g1}
\end{equation}
and the particle-dependent pair-message group summary is
\begin{equation}
\begin{aligned}
    \mathbf{g}^\ell_{2,\alpha}
    =
    \operatorname{concat}
    \left(
        \left\langle \mathbf{m}^\ell_{\alpha I} \right\rangle_{I\in\mathcal{N}},\,
        \left\langle \mathbf{m}^\ell_{\alpha i} \right\rangle_{i\in\mathcal{E}_{\uparrow}},\,
        \left\langle \mathbf{m}^\ell_{\alpha i} \right\rangle_{i\in\mathcal{E}_{\downarrow}}
    \right).
\end{aligned}
\label{eq:app_g2}
\end{equation}

The one-particle update input is
\begin{equation}
    \mathbf{F}^\ell_\alpha
    =
    \operatorname{concat}
    \left(
        \mathbf{h}^\ell_\alpha,\,
        \mathbf{g}^\ell_1,\,
        \mathbf{g}^\ell_{2,\alpha}
    \right).
    \label{eq:app_update_input}
\end{equation}

The one-particle and two-particle representations are then updated as
\begin{equation}
\begin{aligned}
    \mathbf{h}^{\ell+1}_\alpha
    &=
    \frac{1}{\sqrt{2}}
    \Big(
        \mathbf{h}^\ell_\alpha
        +
        \SiLU\big(
            W^\ell_1\mathbf{F}^\ell_\alpha+b^\ell_1
        \big)
    \Big),
    \\
    \mathbf{h}^{\ell+1}_{\alpha\beta}
    &=
    \frac{1}{\sqrt{2}}
    \Big(
        \mathbf{h}^\ell_{\alpha\beta}
        +
        \SiLU\big(
            W^\ell_2\mathbf{h}^\ell_{\alpha\beta}+b^\ell_2
        \big)
    \Big).
\end{aligned}
\label{app_net_backbone_update}
\end{equation}

The layerwise sequence $\mathbf{h}^0_\alpha,\ldots,\mathbf{h}^L_\alpha$ defines the stream associated with particle $\alpha$.
Because electrons and nuclei both carry streams through the shared electron--nucleus backbone, the nuclear representations enter the electronic updates through the group summaries, while the electronic representations analogously enter the nuclear updates.

\subsection{Determinant Construction}
\label{app:determinant_construction}

After the shared electron--nucleus backbone, the final electronic representations are used to construct the orbital matrices and the Jastrow exponent.
The envelope amplitudes and decay parameters are trainable parameters shared across nuclei.

For determinant $k$ and spin channel $\sigma$, let $A^{k\sigma}$ denote the orbital matrix generated from the final electronic representations and let $\Omega^{k\sigma}$ denote the corresponding envelope matrix.
Their elementwise product gives the spin-resolved block
\begin{equation}
    B^{k\sigma}
    =
    A^{k\sigma}\odot\Omega^{k\sigma},
    \qquad
    B^{k\sigma}_{ia}
    =
    A^{k\sigma}_{ia}\Omega^{k\sigma}_{ia},
    \qquad
    i\in\mathcal{E}_{\sigma},
    \quad
    a=1,\ldots,N .
    \label{eq:app_enveloped_orbital_block}
\end{equation}

For each determinant $k$, the spin-resolved blocks are stacked along the electron dimension to form the full Slater matrix,
\begin{equation}
    \Phi_k
    =
    \begin{pmatrix}
        B^{k\uparrow} \\
        B^{k\downarrow}
    \end{pmatrix}
    =
    \begin{pmatrix}
        A^{k\uparrow}\odot\Omega^{k\uparrow} \\
        A^{k\downarrow}\odot\Omega^{k\downarrow}
    \end{pmatrix}
    \in
    \mathbb{R}^{N\times N}.
    \label{eq:app_full_slater_matrix}
\end{equation}

The contribution of determinant $k$ is
\begin{equation}
    D_k
    \left(
        \mathbf{r};\mathbf{R}
    \right)
    =
    \det
    \left[
        \Phi_k
        \left(
            \mathbf{r};\mathbf{R}
        \right)
    \right].
    \label{eq:app_full_determinant}
\end{equation}

The NucleoNet wavefunction is then
\begin{equation}
    \Psi
    \left(
        \mathbf{r};\mathbf{R}
    \right)
    =
    \exp
    \big(
        J(\mathbf{r};\mathbf{R})
    \big)
    \sum_{k=1}^{K}
    D_k
    \left(
        \mathbf{r};\mathbf{R}
    \right).
    \label{eq:app_final_wavefunction}
\end{equation}

Because the rows of the Slater matrix are indexed only by electrons, exchanging two same-spin electrons exchanges the corresponding rows and changes the sign of the determinant.
The nuclear representations do not appear as rows or columns of the Slater matrix and therefore are not included in the fermionic antisymmetrization.
\section{Training and Experimental Details}
\label{app:training_details}

\subsection{Optimizer and VMC Objective}
\label{app:optimizer}

All reported NucleoNet calculations use the SPRING optimizer during the VMC stage.
SPRING~\citep{goldshlager2024spring} is an efficient stochastic-reconfiguration-based optimizer that combines ideas from minimum-step stochastic reconfiguration (MinSR) and the randomized block Kaczmarz method.
At each iteration, Monte Carlo samples define an underdetermined subsampled SR problem through centered log-wavefunction derivatives and local-energy fluctuations.
SPRING updates an approximate SR direction by projecting the previous direction toward the solution of the current subsampled problem, thereby reusing information from previous minibatches.
Tikhonov damping, a decay factor on the previous update direction, and a Euclidean norm constraint are used for stability, while the update is computed in sample space without constructing the full parameter-space SR matrix.

For a wavefunction $\Psi_\theta$, the VMC objective is
\begin{equation}
    E_\theta
    =
    \frac{
        \left\langle
            \Psi_\theta
            \middle|
            \hat H
            \middle|
            \Psi_\theta
        \right\rangle
    }{
        \left\langle
            \Psi_\theta
            \middle|
            \Psi_\theta
        \right\rangle
    }
    =
    \mathbb{E}_{\mathbf{r}\sim|\Psi_\theta|^2}
    \left[
        E_L(\mathbf{r})
    \right],
    \label{eq:app_vmc_objective}
\end{equation}
where the local energy is
\begin{equation}
    E_L(\mathbf{r})
    =
    \frac{
        \hat H\Psi_\theta(\mathbf{r})
    }{
        \Psi_\theta(\mathbf{r})
    } .
    \label{eq:app_local_energy}
\end{equation}

The corresponding VMC gradient can be written as
\begin{equation}
    \nabla_\theta E_\theta
    =
    2
    \mathbb{E}_{\mathbf{r}\sim|\Psi_\theta|^2}
    \left[
        \left(
            E_L(\mathbf{r})-E_\theta
        \right)
        \nabla_\theta
        \log
        \left|
            \Psi_\theta(\mathbf{r})
        \right|
    \right].
    \label{eq:app_vmc_gradient}
\end{equation}

\subsection{Optimizer and Sampling Hyperparameters}
\label{app:optimizer_hyperparameters}

Table~\ref{tab:app_optimizer_hyperparameters} summarizes the main optimizer and sampling hyperparameters. 
All reported runs use the same optimizer hyperparameters unless otherwise stated.
All reported NucleoNet calculations are performed on a single NVIDIA A100 GPU.

\begin{table}[h]
\centering
\caption{
    SPRING and VMC optimization settings.
    Here, $t$ is the zero-based VMC optimization step index and $T_{\mathrm{VMC}}$ is the configured number of VMC optimization steps.
    }
\label{tab:app_optimizer_hyperparameters}
\begin{tabular}{lll}
\hline
Category & Hyperparameter & Value \\
\hline
SPRING
& Learning-rate schedule & $\eta_t=5\times10^{-3}/(1+10^{-4}t)$ \\
& Damping schedule & $\lambda_t=(\lambda_0-\lambda_\infty)/(1+t/c)+\lambda_\infty$ \\
& Initial damping $\lambda_0$ & $10^{-3}$ \\
& Asymptotic damping $\lambda_\infty$ & $10^{-6}$ \\
& Damping decay scale $c$ & $\max(\lfloor T_{\mathrm{VMC}}/100\rfloor,1)$ \\
& History coefficient $\mu$ & $0.99$ \\ 
& Squared update-norm bound & $10^{-3}$ \\
\hline
VMC
& Training iterations & 150000 \\
& Burn-in steps & 4000 \\
& MCMC steps per parameter update & 40 \\
\hline
\end{tabular}

\end{table}

\subsection{Hartree--Fock Pretraining}
\label{app:pretraining}

Hartree--Fock (HF) pretraining is used to initialize the wavefunction before VMC optimization for the systems specified below.
For all equilibrium small-molecule benchmarks, N$_2$ dissociation and LiCN displacement experiments, HF pretraining was used before VMC optimization. 
For the sparse-geometry experiments, HF pretraining uses only the geometries included in the training set.
The reported ethane and formamide calculations use direct VMC optimization without HF pretraining.

Table~\ref{tab:app_pretraining_hyperparameters} summarizes the Hartree--Fock pretraining settings.
\begin{table}[htbp]
    \centering
    \caption{
    Hartree--Fock pretraining settings.
    Pretraining is omitted for the reported ethane and formamide calculations.
    }
    \label{tab:app_pretraining_hyperparameters}
    \renewcommand{\arraystretch}{1.15}
    \begin{tabular}{ll}
    \toprule
    Hyperparameter & Value \\
    \midrule
    Hartree--Fock method & Unrestricted Hartree--Fock \\
    Basis set & cc-pVDZ \\
    Optimizer & Adam \\
    Learning rate & $3\times10^{-4}$ \\
    Pretraining updates & $4000$ \\
    Metropolis steps per parameter update & $20$ \\
    \bottomrule
    \end{tabular}
\end{table}

\subsection{Architecture Hyperparameters}
\label{app:architecture_hyperparameters}

Table~\ref{tab:app_architecture_hyperparameters} summarizes the main architecture hyperparameters. 
The small-molecule benchmarks use a compact NucleoNet configuration, while the multi-geometry benchmarks use a wider configuration.

\begin{table}[h]
\centering
\caption{
Architecture hyperparameters used in the reported calculations.
}
\label{tab:app_architecture_hyperparameters}
\begin{tabular}{lll}
\hline
Category & Hyperparameter & Value \\
\hline
Small molecules
& One-particle width $h_1$ & 64 \\
& Two-particle width $h_2$ & 16 \\
\hline
Multi-geometry systems
& One-particle width $h_1$ & 128 \\
& Two-particle width $h_2$ & 32 \\
\hline
Shared architecture
& Number of electron--nucleus layers & 4 \\
& Number of determinants $K$ & 16 \\
& Orbital type & Real-valued \\
& Activation & $\SiLU$ \\
\hline
Jastrow MLP
& Number of hidden layers & 4 \\
& Hidden width & 32 \\
& Activation & $\tanh$ \\
\hline
\end{tabular}
\end{table}

\subsection{Energy Evaluation and Statistical Uncertainties}
\label{app:energy_evaluation}

After optimization, the wavefunction parameters are held fixed during energy evaluation.
For each nuclear geometry, the energy is estimated by averaging the retained local-energy samples over all sampling chains.
The standard error of the mean energy is corrected for temporal correlations as
\begin{equation}
    \mathrm{SE}(\widehat E)
    =
    \sqrt{
        \frac{\widehat\tau\,\widehat V}
             {n_{\mathrm{sample}}}
    },
    \label{eq:app_energy_standard_error}
\end{equation}
where $\widehat V$ is the estimated local-energy variance, $\widehat\tau$ is the estimated integrated autocorrelation time, and $n_{\mathrm{sample}}$ is the total number of retained local-energy samples for that geometry.

\section{Small-Molecule Benchmark Data}
\label{app:small_molecule_data}

For the equilibrium small-molecule benchmarks, we use a compact NucleoNet configuration with $(h_1,h_2)=(64,16)$ and a batch size of 4096 for all systems. 
Table~\ref{tab:prr_energies} reports the total energies and parameter counts. 
The molecular geometries are taken from the FermiNet benchmark protocol~\citep{pfau2020ferminet}.

\begin{table}[h]
    \scriptsize
    \caption{
        Total energies and parameter counts for eight small molecular systems.
        FermiNet and Psiformer results are taken from~\citep{glehn2023psiformer}, and LapNet results are taken from~\citep{li2024lapnet}. 
        Bold values indicate the lowest reported energy for each system.
        }
    {\centering
    \begin{tabular}{crrrrrrr}
    \hline\hline
    System &  & Psiformer(small) & Psiformer(large) & FermiNet(small) & FermiNet(large) & LapNet & NucleoNet \\
    \hline
       LiH & Energy  &  -8.070528(5) &  -8.070536(4) &  -8.07050(1)  &   -8.070515(8)  & -8.070523(4) &  \textbf{-8.070544(2)} \\
    & \# parameters  &       1.59M &      6.37M         &         0.68M &    2.61M        &        1.88M &         0.10M \\ \hline
    Li$_2$ & Energy &  -14.99486(1) & -14.99485(2)      & -14.99480(2) &    -14.99484(2)   &-14.99485(1) &  \textbf{-14.994913(4)} \\
    & \# parameters  &       1.60M &      6.40M          &  0.70M        &     2.68M        &  1.89M     &         0.10M \\ \hline
    NH$_3$ & Energy  &  -56.56367(2) & -56.56381(2)     & -56.56347(4) &  -56.56387(3)      &-56.56359(2) &  \textbf{-56.56389(1)} \\
    & \# parameters  &       1.62M &      6.47M     &         0.74M      &      2.82M        &1.93M      &        0.11M \\ \hline
    CH$_4$ & Energy  &  -40.51454(2) &  -40.51461(1)    &   -40.51430(3) &  -40.51450(3)    &  -40.51445(2) &  \textbf{-40.51471(1)} \\
    & \# parameters  &       1.62M &      6.47M         &  0.74M         &      2.83M       &   1.93M       &        0.11M \\ \hline
        CO & Energy  & -113.32416(4) & \textbf{-113.32466(3)}    &   -113.32354(7) & -113.32444(5)    &  -113.32417(4) & -113.32465(3) \\
    & \# parameters  &       1.64M &      6.53M         &  0.77M        &      2.94M        &        1.96M &        0.12M \\ \hline
     N$_2$ & Energy  & -109.54137(4) & -109.54179(4)     &  -109.54046(6) & -109.54148(6)   &   -109.54128(4) & \textbf{-109.54195(3)} \\
    & \# parameters   &       1.64M &      6.53M         &  0.77M            &    2.94M       &   1.96M       &        0.12M \\ \hline
    C$_2$H$_4$ & Energy   &  -78.58762(3) &  -78.58794(3)    &-78.58604(5)   &  -78.58701(6)      &-78.58721(4) &  \textbf{-78.58799(3)} \\
    & \# parameters   &       1.65M &      6.58M         &  0.80M        &        3.04M      & 1.99M      &        0.12M \\ \hline
    methylamine & Energy  & -95.86050(4) & \textbf{-95.86096(3)} &  -95.85917(6) & -95.86010(5)   &   -95.86025(3) & -95.86083(4) \\
    & \# parameters   &       1.66M &      6.61M         &  0.82M            &    3.11M       &   2.01M       &        0.13M \\ \hline
    \hline
  \end{tabular}
  \par}

  \label{tab:prr_energies}
\end{table}

\section{Multi-Geometry Benchmark Details}
\label{app:multi_geometry_details}

For the multi-geometry experiments, we use a wider NucleoNet configuration with $(h_1,h_2)=(128,32)$. 
A single set of neural-network parameters is shared across all selected nuclear geometries. 
The optimization objective is the average VMC energy over the training geometries:
\begin{equation}
    \mathcal{L}(\theta)
    =
    \frac{1}{N_{\mathrm{train}}}
    \sum_{s=1}^{N_{\mathrm{train}}}
    \mathbb{E}_{\mathbf{r}\sim|\Psi_\theta(\cdot;\mathbf{R}_s)|^2}
    \left[
        E_L(\mathbf{r};\mathbf{R}_s)
    \right].
    \label{eq:app_multigeometry_objective}
\end{equation}

For the multi-geometry benchmarks, we report signed energy differences relative to the corresponding reference calculations:
\begin{equation}
    \Delta E(s)
    =
    E_{\mathrm{NucleoNet}}(\mathbf{R}_s)
    -
    E_{\mathrm{ref}}(\mathbf{R}_s),
    \label{eq:app_delta_e}
\end{equation}
where $s$ indexes the molecular geometry. 
The mean signed difference is
\begin{equation}
    \mathrm{MSD}
    =
    \frac{1}{N_{\mathrm{geom}}}
    \sum_{s=1}^{N_{\mathrm{geom}}}
    \Delta E(s).
    \label{eq:app_msd}
\end{equation}

When the reference is an empirical or fitted potential energy curve, as in the N$_2$ benchmark, we instead report the mean absolute error:
\begin{equation}
    \mathrm{MAE}
    =
    \frac{1}{N_{\mathrm{geom}}}
    \sum_{s=1}^{N_{\mathrm{geom}}}
    \left|
        E_{\mathrm{NucleoNet}}(\mathbf{R}_s)
        -
        E_{\mathrm{ref}}(\mathbf{R}_s)
    \right|.
    \label{eq:app_mae}
\end{equation}

For ethane and formamide, $E_{\mathrm{ref}}(\mathbf{R}_s)$ denotes the reference calculations reported in Ref.~\citep{foster2025orbformer}.
In these cases, negative signed differences indicate lower optimized variational energies relative to the reported reference calculations.

Table~\ref{tab:app_multi_batchsize} lists the effective batch sizes used for the multi-geometry experiments. 
The notation $n_{\mathrm{chain}}\times N_{\mathrm{geom}}$ denotes the number of Monte Carlo chains per geometry times the number of geometries included in the optimization.

\begin{table}[h]
\centering
\caption{
Batch sizes used in the multi-geometry experiments.
}
\label{tab:app_multi_batchsize}
\begin{tabular}{lll}
\hline
System & Number of geometries & Effective batch size \\
\hline
N$_2$ dissociation & 32 & $128\times 32$ \\
LiCN displacement & 5 & $820\times 5$ \\
Ethane dissociation & 20 & $128\times 20$ \\
Formamide dissociation & 20 & $128\times 20$ \\
\hline
\end{tabular}
\end{table}

\subsection{N$_2$ Dissociation Details}
\label{app:n2_dis}

For the full-geometry N$_2$ dissociation benchmark, a single NucleoNet model is optimized jointly on 32 bond lengths uniformly spaced between $1.6$ and $6.4~a_0$.
The same 32 geometries are used to evaluate the full-geometry result reported in the main text.

The FermiNet and Pfaffian curves shown in the main text are digitized from Ref.~\citep{gao2024neuralpfaffians}, and the PESNet curve is digitized from the original PESNet paper~\citep{gao2022pesnet}.
The explicitly correlated multireference averaged coupled-pair functional ($r_{12}$-MR-ACPF) data are taken directly from the numerical values reported in Ref.~\citep{robert1998r12}.
For each baseline, the MAE is computed over its respective available bond lengths.
Digitization was performed using PlotDigitizer~\citep{plotdigitizer316}.

The reference curve is the MLR$_4(6,8)$ potential of Ref.~\citep{leroy2006accurate}, with its dissociation limit set to $2E_{\mathrm N}$ using $E_{\mathrm N}=-54.5892\,E_h$ from Ref.~\citep{PhysRevA.47.3649}.

The NucleoNet energies on the 32-point grid are reported in Table~\ref{tab:n2_full_energies}.

\begin{table}[t]
    {\centering
    \caption{N$_2$ energies on 32 bond lengths uniformly spaced  between $1.6$ and $6.4~a_0$.}
    \label{tab:n2_full_energies}
    \small
    \setlength{\tabcolsep}{8pt}
    \begin{tabular}{cc@{\hspace{1.2cm}}cc}
    \toprule
    Bond length $L$ ($a_0$) & Energy ($E_h$) &
    Bond length $L$ ($a_0$) & Energy ($E_h$) \\
    \midrule
    1.6000 & $-109.22020(8)$ & 4.0774 & $-109.19612(6)$ \\
    1.7548 & $-109.42544(7)$ & 4.2323 & $-109.19126(6)$ \\
    1.9097 & $-109.51676(7)$ & 4.3871 & $-109.18787(6)$ \\
    2.0645 & $-109.54171(7)$ & 4.5419 & $-109.18552(6)$ \\
    2.2194 & $-109.52889(6)$ & 4.6968 & $-109.18367(6)$ \\
    2.3742 & $-109.49621(6)$ & 4.8516 & $-109.18246(5)$ \\
    2.5290 & $-109.45468(7)$ & 5.0065 & $-109.18157(6)$ \\
    2.6839 & $-109.41088(7)$ & 5.1613 & $-109.18087(5)$ \\
    2.8387 & $-109.36855(7)$ & 5.3161 & $-109.18029(5)$ \\
    2.9935 & $-109.32987(7)$ & 5.4710 & $-109.17982(6)$ \\
    3.1484 & $-109.29610(8)$ & 5.6258 & $-109.17956(6)$ \\
    3.3032 & $-109.26748(7)$ & 5.7806 & $-109.17932(5)$ \\
    3.4581 & $-109.24423(7)$ & 5.9355 & $-109.17902(5)$ \\
    3.6129 & $-109.22625(7)$ & 6.0903 & $-109.17883(5)$ \\
    3.7677 & $-109.21287(8)$ & 6.2452 & $-109.17880(6)$ \\
    3.9226 & $-109.20307(7)$ & 6.4000 & $-109.17857(6)$ \\
    \bottomrule
    \end{tabular}
    \par}
\end{table}

\subsection{LiCN Benchmark Details}
\label{app:licn_details}

For the LiCN benchmark, we consider five collinear configurations from the benchmark used in Orbformer~\citep{foster2025orbformer}.
The Orbformer results correspond to the model fine-tuned from the checkpoint labeled LAC in Ref.~\citep{foster2025orbformer}.
Across these configurations, the C and N nuclei remain fixed, while the Li nucleus is displaced along the molecular axis.
The corresponding Li--C distances, spanning approximately $3.6$ to $4.7~a_0$, are listed in Table~\ref{tab:licn_energies} and define the one-dimensional coordinate used for the Li-displacement benchmark.
The Orbformer and CCSD(T)/CBS energies are taken from Ref.~\citep{foster2025orbformer}.

We perform two sets of NucleoNet calculations.
First, each geometry is optimized independently using a single-geometry NucleoNet model.
These independently optimized energies serve as the baseline for the signed energy differences reported in the main text.
Second, a single NucleoNet model is optimized jointly across all five geometries using shared parameters.
Both the independently optimized single-geometry models and the jointly optimized multi-geometry model use representation widths $(h_1,h_2)=(128,32)$.
The resulting single-geometry and multi-geometry energies are reported in Table~\ref{tab:licn_energies}.

\begin{table}[t]
\caption{
    Li--C distances and NucleoNet energies for the five LiCN configurations.
    The C and N nuclei remain fixed across all configurations, while the Li position varies along the molecular axis.
    The single-geometry energies are obtained from independent optimization at each geometry and serve as the baseline used in the main text.
    The multi-geometry energies are obtained from a single NucleoNet model optimized jointly across all five configurations.
    }
{\centering
\label{tab:licn_energies}
\small
\setlength{\tabcolsep}{7pt}
\begin{tabular}{cccc}
\toprule
Geometry ID &
Li--C distance ($a_0$) &
Single-geometry NucleoNet ($E_h$) &
Multi-geometry NucleoNet ($E_h$) \\
\midrule
1 & 3.590479 & $-100.38046(3)$ & $-100.38045(3)$ \\
2 & 3.779452 & $-100.37951(3)$ & $-100.37945(3)$ \\
3 & 3.968425 & $-100.37624(3)$ & $-100.37625(3)$ \\
4 & 4.346370 & $-100.36622(3)$ & $-100.36614(4)$ \\
5 & 4.724315 & $-100.35389(3)$ & $-100.35385(4)$ \\
\bottomrule
\end{tabular}
\par}
\end{table}

\subsection{Ethane and Formamide Dissociation Details}
\label{app:ethane_formamide_details}

For the ethane and formamide benchmarks, we use the 20-image nudged-elastic-band minimum-energy paths (MEPs) reported in the Orbformer benchmark~\citep{foster2025orbformer}.
The paths describe homolytic C--H cleavage in ethane and N--H cleavage in formamide, with the remaining nuclei allowed to relax along the path.
The reference energies used for the signed energy differences in the main text are independently optimized single-geometry Psiformer VMC energies reported in the same reference.
For each molecule, a single NucleoNet model is optimized jointly across all 20 images.
The resulting NucleoNet total energies are listed in Table~\ref{tab:ethane_formamide_energies}.

\begin{table}[t]
\caption{
NucleoNet energies for the ethane and formamide dissociation benchmarks. 
Image IDs identify the 20 geometries along each dissociation path. 
}
{\centering

\label{tab:ethane_formamide_energies}
\small
\setlength{\tabcolsep}{7pt}
\begin{tabular}{ccc@{\hspace{0.9cm}}ccc}
\toprule
Image ID & Ethane ($E_h$) & Formamide ($E_h$) &
Image ID & Ethane ($E_h$) & Formamide ($E_h$) \\
\midrule
0  & $-79.82448(8)$ & $-169.8992(2)$ &
10 & $-79.64326(7)$ & $-169.8623(2)$ \\

1  & $-79.82254(7)$ & $-169.8995(2)$ &
11 & $-79.64816(7)$ & $-169.8478(2)$ \\

2  & $-79.81594(7)$ & $-169.8957(2)$ &
12 & $-79.64788(7)$ & $-169.7359(2)$ \\

3  & $-79.77018(8)$ & $-169.8853(2)$ &
13 & $-79.64827(7)$ & $-169.6903(2)$ \\

4  & $-79.70948(7)$ & $-169.8650(2)$ &
14 & $-79.64937(7)$ & $-169.6905(2)$ \\

5  & $-79.65531(7)$ & $-169.7275(2)$ &
15 & $-79.65006(7)$ & $-169.6891(2)$ \\

6  & $-79.63373(8)$ & $-169.7114(2)$ &
16 & $-79.6502(1)$  & $-169.6918(2)$ \\

7  & $-79.60163(9)$ & $-169.8353(2)$ &
17 & $-79.65056(7)$ & $-169.6916(2)$ \\

8  & $-79.59682(9)$ & $-169.8592(2)$ &
18 & $-79.65056(7)$ & $-169.6911(3)$ \\

9  & $-79.63426(8)$ & $-169.8621(2)$ &
19 & $-79.65038(7)$ & $-169.6913(2)$ \\
\bottomrule
\end{tabular}
\par}
\end{table}

\section{Generalization from Sparse Geometry Sets}
\label{app:geometry_transfer}

We evaluate generalization by optimizing NucleoNet models on sparse sets of training geometries and evaluating them on denser geometry grids, most of whose geometries are not included in the optimization.
These experiments are performed for N$_2$ and LiCN.
No additional wavefunction optimization is performed at the unseen geometries.

Table~\ref{tab:app_transfer_batchsize} summarizes the effective batch sizes used in the sparse-geometry experiments.
The notation $n_{\mathrm{chain}}\times N_{\mathrm{opt}}$ denotes the number of Monte Carlo chains per optimization geometry multiplied by the number of geometries included in the optimization.

\begin{table}[h]
\centering
\caption{
Batch sizes used in the sparse-geometry experiments.
}
\label{tab:app_transfer_batchsize}
\begin{tabular}{lll}
\toprule
System & Optimization geometries & Effective batch size \\
\midrule
N$_2$ & 6 & $684\times 6$ \\
LiCN  & 2 & $2048\times 2$ \\
\bottomrule
\end{tabular}
\end{table}

\subsection{N$_2$ Generalization}
\label{app:n2_transfer}

For the N$_2$ sparse-geometry experiment, a separate NucleoNet model is optimized on six training geometries, bond lengths uniformly spaced between $1.6$ and $6.0~a_0$.
After optimization, the model is evaluated without further parameter updates on a 32-point bond-length grid uniformly spaced between $1.6$ and $6.4~a_0$.

The sparse-geometry result is compared with the full-geometry joint-optimization result described in Sec.~\ref{app:n2_dis}.
The six optimization geometries and corresponding energies are listed in Table~\ref{tab:n2_training_grid}.

\begin{table}[t]
\centering
\caption{
N$_2$ optimization geometries and corresponding energies for the sparse-geometry experiment.
}
\label{tab:n2_training_grid}
\small
\setlength{\tabcolsep}{10pt}
\begin{tabular}{cc}
\toprule
Bond length $L$ ($a_0$) & Energy ($E_h$) \\
\midrule
1.6000 & $-109.22046(5)$ \\
2.4800 & $-109.46837(5)$ \\
3.3600 & $-109.25831(5)$ \\
4.2400 & $-109.19088(5)$ \\
5.1200 & $-109.18093(4)$ \\
6.0000 & $-109.17897(4)$ \\
\bottomrule
\end{tabular}
\end{table}

The energies obtained on the complete evaluation grid are reported in Table~\ref{tab:n2_transfer_energies}.

\begin{table}[t]
    {\centering
    \caption{
    N$_2$ energies on a 32-point bond-length grid uniformly spaced between $1.6$ and $6.4~a_0$ for the sparse-geometry experiment.
    }
    \label{tab:n2_transfer_energies}
    \small
    \setlength{\tabcolsep}{8pt}
    \begin{tabular}{cc@{\hspace{1.2cm}}cc}
    \toprule
    Bond length $L$ ($a_0$) & Energy ($E_h$) &
    Bond length $L$ ($a_0$) & Energy ($E_h$) \\
    \midrule
    1.6000 & $-109.22032(7)$ & 4.0774 & $-109.19600(6)$ \\
    1.7548 & $-109.42535(7)$ & 4.2323 & $-109.19123(7)$ \\
    1.9097 & $-109.51670(7)$ & 4.3871 & $-109.18768(6)$ \\
    2.0645 & $-109.54159(8)$ & 4.5419 & $-109.18527(6)$ \\
    2.2194 & $-109.52875(7)$ & 4.6968 & $-109.18351(7)$ \\
    2.3742 & $-109.49613(7)$ & 4.8516 & $-109.18239(6)$ \\
    2.5290 & $-109.45462(7)$ & 5.0065 & $-109.18147(6)$ \\
    2.6839 & $-109.41081(7)$ & 5.1613 & $-109.18079(5)$ \\
    2.8387 & $-109.36852(7)$ & 5.3161 & $-109.18030(6)$ \\
    2.9935 & $-109.32978(7)$ & 5.4710 & $-109.17991(6)$ \\
    3.1484 & $-109.29595(7)$ & 5.6258 & $-109.17958(5)$ \\
    3.3032 & $-109.26746(7)$ & 5.7806 & $-109.17929(6)$ \\
    3.4581 & $-109.24428(7)$ & 5.9355 & $-109.17910(6)$ \\
    3.6129 & $-109.22619(7)$ & 6.0903 & $-109.17899(6)$ \\
    3.7677 & $-109.21268(7)$ & 6.2452 & $-109.17863(8)$ \\
    3.9226 & $-109.20297(7)$ & 6.4000 & $-109.17858(7)$ \\
    \bottomrule
    \end{tabular}
    \par}
\end{table}

\subsection{LiCN Generalization}
\label{app:licn_transfer}

For the LiCN sparse-geometry experiment, a separate NucleoNet model is optimized only on the two endpoint configurations as training geometries (Geometry IDs 1 and 5).
The optimized model is then evaluated directly on the three intermediate configurations (Geometry IDs 2--4) without further wavefunction optimization.

The resulting five-geometry energy profile is compared with the full-geometry NucleoNet model optimized jointly on all five configurations in Sec.~\ref{app:licn_details}.
Both calculations are evaluated relative to the independently optimized single-geometry NucleoNet energies used as the baseline in the main text.

The energies evaluated with the sparsely optimized model are listed in Table~\ref{tab:licn_transfer_energies}.

\begin{table}[t]
\caption{
    LiCN energies obtained with the model optimized on the two endpoint geometries.
    All five configurations are evaluated without further wavefunction optimization.
}
{\centering
\label{tab:licn_transfer_energies}
\small
\setlength{\tabcolsep}{8pt}
\begin{tabular}{ccc}
\toprule
Geometry ID & Role & Sparse-geometry energy ($E_h$) \\
\midrule
1 & Optimization & $-100.38046(3)$ \\
2 & Evaluation   & $-100.37944(4)$ \\
3 & Evaluation   & $-100.37626(4)$ \\
4 & Evaluation   & $-100.36622(3)$ \\
5 & Optimization & $-100.35378(7)$ \\
\bottomrule
\end{tabular}
\par}
\end{table}

\section{Ablation Details}
\label{app:ablation}

This section defines the four models used to examine envelope parameter sharing and nuclear propagation.
The suffix ``-NS'' denotes nucleus-specific envelope coefficients, and ``-EB'' denotes an electron-only backbone.
Their combinations are summarized in Table~\ref{tab:app_ablation_variants}.
The single-geometry and sparse-geometry ablations use the architecture widths and training settings of the corresponding benchmarks.
All four variants use the same number of determinants, backbone depth, and optimization budget within each benchmark.

\begin{table}[t]
\centering
\caption{
Backbone and envelope choices for the four ablation models.
Shared coefficients are common to all nuclei, whereas nucleus-specific coefficients are optimized separately for each nucleus.
}
\label{tab:app_ablation_variants}
\begin{tabular}{lll}
\toprule
Model & Backbone & Envelope coefficients \\
\midrule
NucleoNet & Electron--nucleus & Shared \\
NucleoNet-NS & Electron--nucleus & Nucleus-specific \\
NucleoNet-EB & Electron-only & Shared \\
NucleoNet-EB-NS & Electron-only & Nucleus-specific \\
\bottomrule
\end{tabular}
\end{table}

\subsection{Envelope Parameterizations}
\label{app:ablation_envelope}

All four models use isotropic exponential envelopes.
For NucleoNet and NucleoNet-EB, the coefficients are shared across nuclei:
\begin{equation}
\Omega^{k\sigma}_{ia}(\mathbf r_i;\mathbf R)
=
\pi^{k\sigma}_{a}
\sum_{I\in\mathcal N}
\exp\left(
-\lambda^{k\sigma}_{a}
\left\|\mathbf r_i-\mathbf R_I\right\|
\right).
\label{eq:app_ablation_shared_envelope}
\end{equation}
Here, $i\in\mathcal E_\sigma$, $a$ indexes orbitals, and $k$ indexes determinants.
The trainable coefficients $\pi^{k\sigma}_{a}$ and $\lambda^{k\sigma}_{a}$ do not depend on the nuclear index $I$.

For NucleoNet-NS and NucleoNet-EB-NS, each nucleus has separate coefficients:
\begin{equation}
\Omega^{k\sigma}_{ia}(\mathbf r_i;\mathbf R)
=
\sum_{I\in\mathcal N}
\pi^{k\sigma}_{Ia}
\exp\left(
-\lambda^{k\sigma}_{Ia}
\left\|\mathbf r_i-\mathbf R_I\right\|
\right).
\label{eq:app_ablation_ns_envelope}
\end{equation}
This is the nucleus-specific isotropic envelope used in the simplified FermiNet architecture~\citep{spencer2020better}.
In both parameterizations, the envelope coefficients are independent of the nuclear representations.
The envelope depends on nuclear geometry through the electron--nucleus distances.

\subsection{Electron-only Backbone}
\label{app:ablation_fe_eb}

NucleoNet and NucleoNet-NS propagate electronic and nuclear representations through the joint backbone.
NucleoNet-EB and NucleoNet-EB-NS instead propagate only electronic one-particle streams and electron--electron pair representations.
Their propagated particle set is
\begin{equation}
\mathcal P_{\mathrm{EB}}
=
\mathcal E_\uparrow\cup\mathcal E_\downarrow,
\label{eq:app_ablation_eb_particle_set}
\end{equation}
rather than the full set $\mathcal P=\mathcal E_\uparrow\cup\mathcal E_\downarrow\cup\mathcal N$.

Nuclear geometry remains available through the initialization features and the nuclear-centered envelope.
The initial one-particle feature of electron $i$ averages its pair features over all electrons and nuclei:
\begin{equation}
\mathbf u_i^{\mathrm{EB}}
=
\left\langle\mathbf u_{i\alpha}\right\rangle_{\alpha\in\mathcal P},
\qquad i\in\mathcal P_{\mathrm{EB}}.
\label{eq:app_ablation_eb_initial_feature}
\end{equation}
The initial electronic representations are
\begin{equation}
\begin{aligned}
\mathbf h_i^0
&=
\SiLU\left(
W_1^0\operatorname{concat}
\left(\mathbf u_i^{\mathrm{EB}},\mathbf c_i\right)+b_1^0
\right),\\
\mathbf h_{ij}^0
&=
\SiLU\left(
W_2^0\operatorname{concat}
\left(\mathbf u_{ij},\mathbf c_{ij}\right)+b_2^0
\right),
\end{aligned}
\label{eq:app_ablation_eb_initial_representations}
\end{equation}
where $i,j\in\mathcal P_{\mathrm{EB}}$.

At layer $\ell$, the pair-gated message between electrons is
\begin{equation}
\mathbf m_{ij}^{\ell}
=
\mathbf h_{ij}^{\ell}\odot\hat{\mathbf h}_j^{\ell},
\label{eq:app_ablation_eb_pair_message}
\end{equation}
where $\hat{\mathbf h}_j^\ell$ is the projected one-particle representation defined for the full backbone.
The group summaries include only the two electronic spin channels:
\begin{equation}
\mathbf g_{1,\mathrm{EB}}^\ell
=
\operatorname{concat}\left(
\left\langle\mathbf h_j^\ell\right\rangle_{j\in\mathcal E_\uparrow},
\left\langle\mathbf h_j^\ell\right\rangle_{j\in\mathcal E_\downarrow}
\right),
\label{eq:app_ablation_eb_g1}
\end{equation}
\begin{equation}
\mathbf g_{2,i,\mathrm{EB}}^\ell
=
\operatorname{concat}\left(
\left\langle\mathbf m_{ij}^\ell\right\rangle_{j\in\mathcal E_\uparrow},
\left\langle\mathbf m_{ij}^\ell\right\rangle_{j\in\mathcal E_\downarrow}
\right).
\label{eq:app_ablation_eb_g2}
\end{equation}
The one-particle update input is
\begin{equation}
\mathbf F_{i,\mathrm{EB}}^\ell
=
\operatorname{concat}\left(
\mathbf h_i^\ell,
\mathbf g_{1,\mathrm{EB}}^\ell,
\mathbf g_{2,i,\mathrm{EB}}^\ell
\right).
\label{eq:app_ablation_eb_update_input}
\end{equation}
The electronic representations follow the same form of residual updates as in the full model:
\begin{equation}
\begin{aligned}
\mathbf h_i^{\ell+1}
&=
\frac{1}{\sqrt{2}}
\left[
\mathbf h_i^\ell+
\SiLU\left(W_1^\ell\mathbf F_{i,\mathrm{EB}}^\ell+b_1^\ell\right)
\right],\\
\mathbf h_{ij}^{\ell+1}
&=
\frac{1}{\sqrt{2}}
\left[
\mathbf h_{ij}^\ell+
\SiLU\left(W_2^\ell\mathbf h_{ij}^\ell+b_2^\ell\right)
\right].
\end{aligned}
\label{eq:app_ablation_eb_updates}
\end{equation}
The final electronic representations are used to construct the orbital matrices and the Jastrow factor as in NucleoNet.

\subsection{Evaluation Metrics}
\label{app:ablation_metrics}

For the single-geometry benchmarks, the energy error of model $M$ is
\begin{equation}
\Delta E(M)=E_M-E_{\mathrm{ref}},
\label{eq:app_ablation_delta_energy}
\end{equation}
where $E_{\mathrm{ref}}$ is the CCSD(T)/CBS reference energy used in the corresponding benchmark.

For the N$_2$ sparse-geometry benchmark, the mean absolute error on a geometry set $\mathcal S$ is
\begin{equation}
\mathrm{MAE}_{\mathcal S}(M)
=
\frac{1}{|\mathcal S|}
\sum_{s\in\mathcal S}
\left|
E_M(\mathbf R_s)-E_{\mathrm{ref}}(\mathbf R_s)
\right|.
\label{eq:app_ablation_mae}
\end{equation}
We report this metric separately for the six training geometries, $\mathcal S_{\mathrm{train}}$, and the 32-point evaluation grid, $\mathcal S_{\mathrm{eval}}$.
All four models use the same dissociation-limit-anchored MLR$_4(6,8)$ reference described in Sec.~\ref{app:n2_dis}.

\subsection{Raw Ablation Data}
\label{app:ablation_raw_data}

Tables~\ref{tab:app_ablation_single_raw}, \ref{tab:app_ablation_n2_train_raw}, and \ref{tab:app_ablation_n2_eval_raw} report the absolute variational energies for the single-geometry benchmarks and the N$_2$ training and evaluation grids, respectively.
All energies are given in $E_h$, with statistical uncertainties in parentheses.

\begin{table*}[t]
\centering
\caption{
Variational energies for the NH$_3$ and N$_2$ single-geometry ablation benchmarks.
All energies are reported in $E_h$.
}
\label{tab:app_ablation_single_raw}
\small
\setlength{\tabcolsep}{8pt}
\begin{tabular}{lcccc}
\toprule
System & NucleoNet & NucleoNet-NS & NucleoNet-EB & NucleoNet-EB-NS \\
\midrule
NH$_3$ & -56.56389(1) & -56.56388(2) & -56.56384(2) & -56.56384(2) \\
N$_2$ & -109.54195(3) & -109.54184(3)& -109.54177(4) & -109.54173(4) \\
\bottomrule
\end{tabular}
\end{table*}

\begin{table}[t]
\centering
\caption{
N$_2$ energies at the six training geometries for the four sparse-geometry ablation models.
All energies are reported in $E_h$, with standard errors in parentheses.
}
\label{tab:app_ablation_n2_train_raw}
\small
\setlength{\tabcolsep}{8pt}
\begin{tabular}{ccccc}
\toprule
Bond length $L$ ($a_0$) & NucleoNet & NucleoNet-NS & NucleoNet-EB & NucleoNet-EB-NS \\
\midrule
1.6000 & $-109.22046(5)$ & $-109.22033(7)$ & $-109.22001(9)$ & $-109.21963(9)$ \\
2.4800 & $-109.46837(5)$ & $-109.46817(7)$ & $-109.46795(8)$ & $-109.46775(9)$ \\
3.3600 & $-109.25831(5)$ & $-109.25818(8)$ & $-109.25804(9)$ & $-109.25729(9)$ \\
4.2400 & $-109.19088(5)$ & $-109.19091(7)$ & $-109.19074(8)$ & $-109.19006(9)$ \\
5.1200 & $-109.18093(4)$ & $-109.18032(6)$ & $-109.18089(7)$ & $-109.17995(8)$ \\
6.0000 & $-109.17897(4)$ & $-109.17877(5)$ & $-109.17882(7)$ & $-109.1777(1)$ \\
\bottomrule
\end{tabular}
\end{table}

\begin{table}[t]
\centering
\caption{
N$_2$ energies on the 32-point evaluation grid for the four sparse-geometry ablation models.
All energies are reported in $E_h$, with standard errors in parentheses.
}
\label{tab:app_ablation_n2_eval_raw}
\small
\setlength{\tabcolsep}{8pt}
\begin{tabular}{ccccc}
\toprule
Bond length $L$ ($a_0$) & NucleoNet & NucleoNet-NS & NucleoNet-EB & NucleoNet-EB-NS \\
\midrule
1.6000 & $-109.22032(7)$ & $-109.22020(7)$ & $-109.22017(9)$ & $-109.2197(1)$ \\
1.7548 & $-109.42535(7)$ & $-109.42536(7)$ & $-109.4249(1)$ & $-109.4011(5)$ \\
1.9097 & $-109.51670(7)$ & $-109.51666(7)$ & $-109.5161(1)$ & $-108.997(3)$ \\
2.0645 & $-109.54159(8)$ & $-109.54156(7)$ & $-109.5410(1)$ & $-108.896(4)$ \\
2.2194 & $-109.52875(7)$ & $-109.52881(7)$ & $-109.52833(9)$ & $-109.5015(8)$ \\
2.3742 & $-109.49613(7)$ & $-109.49617(7)$ & $-109.49588(9)$ & $-109.4950(2)$ \\
2.5290 & $-109.45462(7)$ & $-109.45462(7)$ & $-109.45414(8)$ & $-109.4538(1)$ \\
2.6839 & $-109.41081(7)$ & $-109.41075(7)$ & $-109.41043(8)$ & $-109.4100(1)$ \\
2.8387 & $-109.36852(7)$ & $-109.36845(7)$ & $-109.36802(9)$ & $-109.3677(1)$ \\
2.9935 & $-109.32978(7)$ & $-109.32976(7)$ & $-109.3296(1)$ & $-109.3291(1)$ \\
3.1484 & $-109.29595(7)$ & $-109.29575(8)$ & $-109.29551(9)$ & $-109.2951(1)$ \\
3.3032 & $-109.26746(7)$ & $-109.26731(7)$ & $-109.26694(9)$ & $-109.2665(1)$ \\
3.4581 & $-109.24428(7)$ & $-109.24419(8)$ & $-109.24408(9)$ & $-109.2434(1)$ \\
3.6129 & $-109.22619(7)$ & $-109.22614(7)$ & $-109.22586(9)$ & $-109.22538(9)$ \\
3.7677 & $-109.21268(7)$ & $-109.21267(7)$ & $-109.21257(8)$ & $-109.2115(1)$ \\
3.9226 & $-109.20297(7)$ & $-109.20284(7)$ & $-109.20263(9)$ & $-109.20210(9)$ \\
4.0774 & $-109.19600(6)$ & $-109.19577(8)$ & $-109.19578(9)$ & $-109.19494(9)$ \\
4.2323 & $-109.19123(7)$ & $-109.1907(2)$ & $-109.19097(8)$ & $-109.19014(9)$ \\
4.3871 & $-109.18768(6)$ & $-109.18758(7)$ & $-109.18765(8)$ & $-109.18666(9)$ \\
4.5419 & $-109.18527(6)$ & $-109.18498(8)$ & $-109.18528(7)$ & $-109.18427(8)$ \\
4.6968 & $-109.18351(7)$ & $-109.18325(7)$ & $-109.18351(7)$ & $-109.18266(8)$ \\
4.8516 & $-109.18239(6)$ & $-109.18203(7)$ & $-109.18230(7)$ & $-109.18141(9)$ \\
5.0065 & $-109.18147(6)$ & $-109.18094(6)$ & $-109.18136(7)$ & $-109.18054(9)$ \\
5.1613 & $-109.18079(5)$ & $-109.1792(4)$ & $-109.18071(7)$ & $-109.17979(8)$ \\
5.3161 & $-109.18030(6)$ & $-109.17997(6)$ & $-109.18017(7)$ & $-109.17941(9)$ \\
5.4710 & $-109.17991(6)$ & $-109.17944(6)$ & $-109.17965(7)$ & $-109.17892(8)$ \\
5.6258 & $-109.17958(5)$ & $-109.17922(6)$ & $-109.17935(7)$ & $-109.1784(1)$ \\
5.7806 & $-109.17929(6)$ & $-109.17897(6)$ & $-109.17918(7)$ & $-109.1782(1)$ \\
5.9355 & $-109.17910(6)$ & $-109.17892(6)$ & $-109.17894(8)$ & $-109.1782(1)$ \\
6.0903 & $-109.17899(6)$ & $-109.17862(6)$ & $-109.17869(7)$ & $-109.1779(1)$ \\
6.2452 & $-109.17863(8)$ & $-109.17844(6)$ & $-109.17860(7)$ & $-109.17765(9)$ \\
6.4000 & $-109.17858(7)$ & $-109.17846(6)$ & $-109.17847(8)$ & $-109.1775(1)$ \\
\bottomrule
\end{tabular}
\end{table}

\clearpage
\bibliography{references}